\documentclass[pdflatex,sn-mathphys-num]{sn-jnl}

\usepackage{graphicx}%
\usepackage{multirow}%
\usepackage{amsmath,amssymb,amsfonts}%
\usepackage{amsthm}%
\usepackage{mathrsfs}%
\usepackage[title]{appendix}%
\usepackage{xcolor}%
\usepackage{textcomp}%
\usepackage{manyfoot}%
\usepackage{booktabs}%
\usepackage{algorithm}%
\usepackage{algorithmicx}%
\usepackage{algpseudocode}%
\usepackage{soul}
\usepackage{listings}%
\usepackage[utf8]{inputenc} 
\usepackage[T1]{fontenc}    
\usepackage{siunitx}
\usepackage{bm}

\theoremstyle{thmstyleone}%
\theoremstyle{thmstyletwo}%

\theoremstyle{thmstylethree}%

\begin{document}

\title[Article Title]{Electrically Reconfigurable Extended Lasing State in an Organic Liquid-Crystal Microcavity}


\author*[1]{\fnm{Dmitriy} \sur{Dovzhenko}}\email{dovzhenkods@gmail.com}
\author[2,3]{\fnm{Luciano} \sur{Siliano Ricco}}
\author[1,4]{\fnm{Krzysztof} \sur{Sawicki}}
\author[3]{\fnm{Marcin} \sur{Muszy\'nski}}
\author[5]{\fnm{Pavel} \sur{Kokhanchik}}
\author[3]{\fnm{Piotr} \sur{Kapu\'sci\'nski}}
\author[6]{\fnm{Przemys\l{}aw} \sur{Morawiak}}
\author[6]{\fnm{Wiktor} \sur{Piecek}}
\author[7]{\fnm{Piotr} \sur{Nyga}}
\author[8]{\fnm{Przemys\l{}aw} \sur{Kula}}
\author[5,9]{\fnm{Dmitry} \sur{Solnyshkov}}
\author[5]{\fnm{Guillaume} \sur{Malpuech}}
\author[3]{\fnm{Helgi} \sur{Sigurðsson}}
\author[3]{\fnm{Jacek} \sur{Szczytko}}
\author[1,10]{\fnm{Simone} \sur{De Liberato}}

\affil[1]{\orgdiv{School of Physics and Astronomy}, \orgname{University of Southampton}, \orgaddress{\street{University Road}, \city{Southampton}, \postcode{SO17 1BJ}, \country{United Kingdom}}}

\affil[2]{\orgdiv{Science Institute}, \orgname{University of Iceland}, \orgaddress{\street{Dunhagi-3}, \city{Reykjavik}, \postcode{IS-107}, \country{Iceland}}}

\affil[3]{\orgdiv{Institute of Experimental Physics, Faculty of Physics}, \orgname{University of Warsaw}, \orgaddress{\street{ulica Pasteura 5}, \city{Warsaw}, \postcode{PL-02-093}, \country{Poland}}}

\affil[4]{\orgdiv{Department of Physics}, \orgname{Durham University}, \orgaddress{\street{South Road}, \city{Durham}, \postcode{DH1 3LE}, \country{United Kingdom}}}

\affil[5]{\orgdiv{Institut Pascal}, \orgname{Universit\'e Clermont Auvergne, CNRS}, \orgaddress{\street{ClermontINP}, \city{Clermont-Ferrand}, \postcode{F-63000}, \country{France}}}

\affil[6]{\orgdiv{Institute of Applied Physics}, \orgname{Military University of Technology}, \orgaddress{\street{S. Kaliskiego 2}, \city{Warsaw}, \postcode{00-908}, \country{Poland}}}

\affil[7]{\orgdiv{Institute of Optoelectronics}, \orgname{Military University of Technology}, \orgaddress{\street{S. Kaliskiego 2}, \city{Warsaw}, \postcode{00-908}, \country{Poland}}}

\affil[8]{\orgdiv{Institute of Chemistry}, \orgname{Military University of Technology}, \orgaddress{\street{S. Kaliskiego 2}, \city{Warsaw}, \postcode{00-908}, \country{Poland}}}

\affil[9]{\orgdiv{Institut Universitaire de France},  \orgaddress{\city{Paris}, \postcode{F-75231}, \country{France}}}

\affil[10]{\orgdiv{Istituto di Fotonica e Nanotecnologie}, \orgname{Consiglio Nazionale delle Ricerche (CNR)}, \orgaddress{\street{Piazza Leonardo da Vinci 32}, \city{Milano}, \postcode{20133}, \country{Italy}}}

\abstract{Small-footprint, low-power, and reprogrammable arrays of coupled coherent emitters are highly sought in modern nanophotonics. Among existing solutions, only inorganic semiconductor microcavities operating in a strong light-matter coupling regime exhibit controlled on-chip interaction between individual coherent states, predominantly at cryogenic temperatures. Here we demonstrate electrically controlled in-plane interaction between optically reconfigurable spatially separated lasing states, operating at room temperature in the weak light-matter coupling regime. An organic liquid crystal-filled microcavity is introduced as a new material platform where a spatially extended coherent lasing state, or "supermode", appears due to the blueshift-induced near-field transverse coupling between distinct spatially pumped states. We demonstrate a wide-range microscale control of supermode near- and far-field with on-chip phase-locking tuning functionality. We realize electrical control over the interaction strength between lasing states and corresponding mutual coherence going beyond nearest neighbours, and a spin-selective directional coupling regime by using a photonic analogue of the Rashba-Dresselhaus spin-orbit interaction.
}


\maketitle

\section{Introduction}\label{sec1}

Creating compact and reprogrammable sources of coherent emission is a central task for developing modern integrated photonic devices. Of separate interest is the realisation of phase-locking in arrays of coherently emitting lasers for beam shaping~\cite{Forbes2021}, topological optics~\cite{Piccardo2022, Dai2024}, investigation of many-body phenomena with optical simulations~\cite{Parto2020, berloff2017realizing} and in the emerging fields of all-optical neural networks~\cite{Chen2023,Matuszewski_PhysRevApplied.21.014028_2024} and reservoir computing~\cite{TANAKA2019100,BallariniACS_nanolett0c004352020}, where optical networks are beneficial due to the high clock rates, parallelism and low losses.

Several approaches have been developed to achieve the phase-locked arrays of lasers, including shaping bulk laser beams with diffractive elements, free-form optics, metasurface arrays, or spatial light modulators~\cite{Ngcobo2013, KumarReddy2022, Piccardo2022}, and phase-locking of the vertical-cavity surface-emitting laser (VCSEL) arrays~\cite{Ma2022, Pan2024}.
Despite recent technological advances, existing solutions usually require either external devices to provide optical feedback and phase injection locking or utilise irreversible fabrication techniques, such as beam shaping using metasurface-integrated VCSELs~\cite{Xie2020} and photonic crystal surface-emitting lasers~\cite{Miyai2006, Raftery2005, Noda2024}.

An alternative approach to compact on-chip optical devices with good in-situ tuneability over the emission characteristics is realised with optical lattices of non-equilibrium exciton-polariton Bose-Einstein condensates~\cite{kasprzak2006bose, Topfer2021} which, due to their strong nonlinearities and optically tunable ballistic coupling mechanism~\cite{Alyatkin2020, Sawicki:26}, are being explored as promising reconfigurable elements for all-optical computing~\cite{Kavokin2022, Opala_OptMatExpress_2023}, spin simulation~\cite{berloff2017realizing, Tao2022}, angular momentum generation~\cite{Cookson2021, Werner2025}, topologically protected lasing~\cite{Pickup2020, Pieczarka2021, Betzold2024}, and more. Although many of the aforementioned works have been conducted at cryogenic temperatures in II-VI or III-V semiconductor quantum well microcavities, increasing effort has been shifted towards more application-friendly room temperature materials~\cite{Sanvitto2016, Ghosh2022, Zasedatelev2021, Shi2025}, in particular organic materials relevant to the current study~\cite{Plumhof2014, Zasedatelev2019, Dusel2020}. However, for room temperature organic-based light-emission devices, the coherent emission usually appears perpendicular to the cavity surface (at $k_{||}=0$) and therefore coupling between the emission nodes (i.e., pumped condensates) is hard to achieve and usually requires significant overlap between the condensate wave-functions~\cite{Plumhof2014, Zasedatelev2019, McGhee2022, Yadav2024, Deshmukh2024}. Ballistic in-plane coupling between coherent emitters at room temperature was though demonstrated with perovskite-based microcavities, where two-dimensional condensate lattices were realized for simulation of XY Hamiltonian \cite{Tao2022, Peng2024}. It is important to note that this approach requires demanding technology to grow large and uniform single-crystal perovskites inside the microcavity, which needs to be simultaneously robust towards photobleaching for continuous operation, limiting the applicability. Besides, the requirement for strong light-matter coupling imposes strict conditions on the parameters of both photonic and excitonic components of the system, restricting tuneability and scalability. In contrast, weakly-coupled organic microcavities offer application-friendly, well-developed, and relatively easy fabrication technology, robustness of operation, and wide range electrical tuning capabilities enabled by the integration with highly birefringent liquid crystals\cite{Muszynski2022}. In this work we attempt to fill the gap between the optically reconfigurable ballistic lattices of polaritonic condensates and tuneable organic microcavities operating at room temperature in the weak light-matter coupling regime, where in-plane phase-locking between separated optically excited lasing states has yet not been reported.

\begin{figure}[!h]
    \centering
    \includegraphics[width=0.95\linewidth]{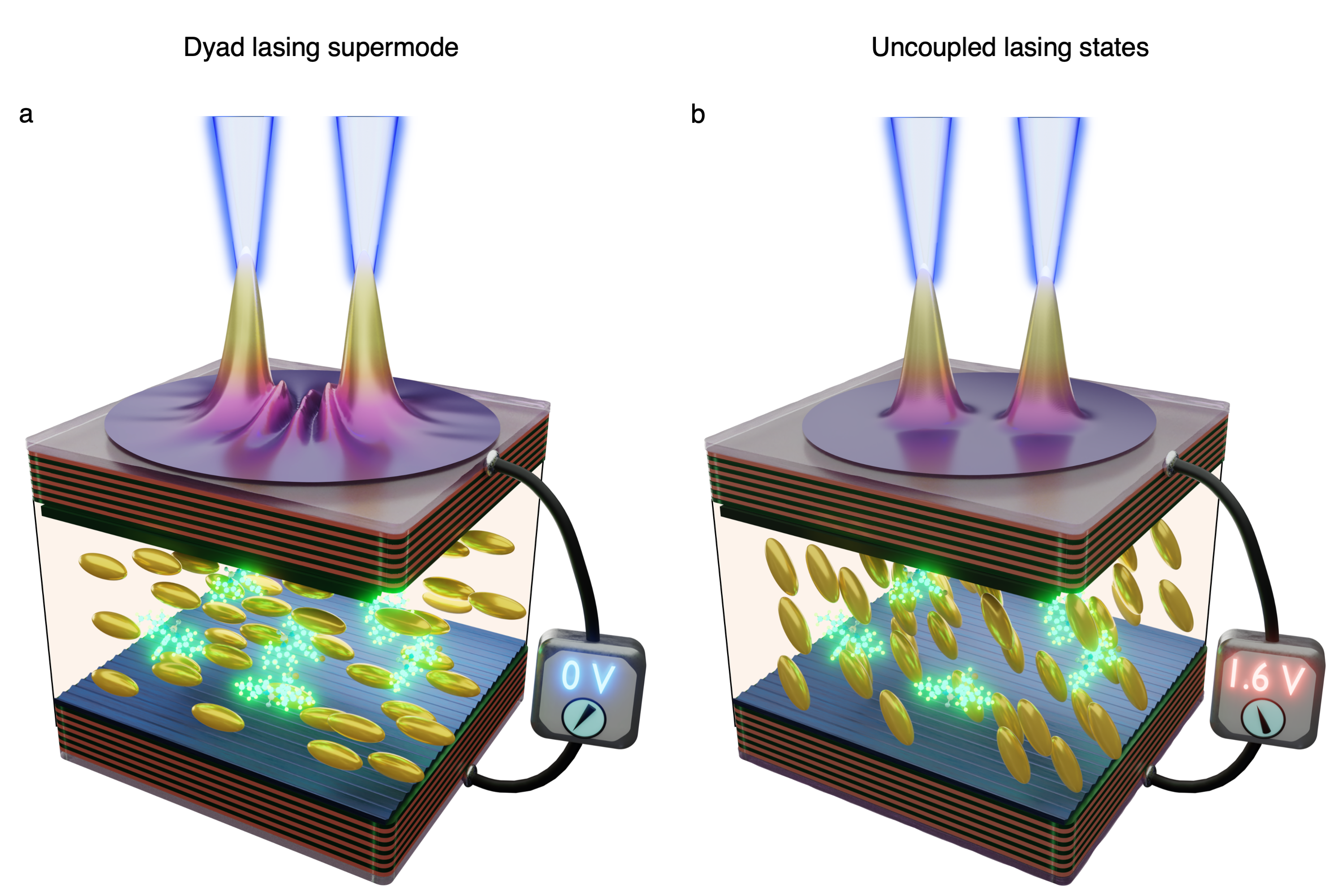}
    \caption{{\bf Electrically controlled dyad lasing supermode}. Schematic showing the microcavity filled with liquid crystals and P580 laser dye pumped in a dyad configuration with {\bf (a)} phase-locked spatially extended supermode without external voltage applied and {\bf (b)} two uncoupled lasing states at 1600 mV voltage correspondingly.}
    \label{fig:schematic}
\end{figure}

Here we present a realisation of spatially extended lasing states, or ``supermodes``, formed by coherent in-plane coupling between individually pumped lasing states with all-optical control over the geometry of phase-locked lasing domains at room temperature in the weak light-matter coupling regime. We demonstrate electrical control over mutual coupling between the lasing states allowing for a wide-range shaping of the near- and far-field of the arrays of coherently coupled emitters, see Fig.~\ref{fig:schematic}(a) for schematic. We utilize a liquid crystal (LC) dye-filled microcavity (LCMC) in the weak light-matter coupling regime, with the photonic wavefunction of the supermode directly accessible by measuring the emission intensity in real space. From here on, a single ``lasing state'' refers to a spatially localized emission site (node) in the microcavity plane that is being optically pumped above threshold with a corresponding tightly focused, normal incidence, nonresonant Gaussian beam. The term ``supermode``~[\cite{Paoli1984, Mehuys1998, Ginis2023}] is used here to refer to the spatially extended coherent lasing state consisting of two or more phase-locked spatially separated and individually pumped lasing spots, which are coupled via in-plane coherent photon exchange. By optically reconfiguring the profile of the incident nonresonant excitation beam into multiple spots, one can easily control the pumping geometry and arbitrary shape of the resulting extended lasing state. The ballistic near field coupling, and subsequent synchronization, between individual lasing states in the plane of the cavity is defined by their in-plane photon momentum component which determines their mutual overlap between the pumped gain regions. Hence, the coupling can be controlled electrically by applying a voltage to the LC-filled microcavity, see Fig.~\ref{fig:schematic}(a,b), which rotates the birefringent molecules, changing the dielectric cavity tensor, and subsequently the in-plane dispersion relation of the cavity photons. At higher voltages, where the transverse electric (TE) mode is tuned into resonance with the next order transverse magnetic (TM) mode of opposite parity~\cite{Rechcinska_Science2019} an effective Rashba-Dresselhaus (RD) spin-orbit coupling (SOC) of the photon dispersion emerges ~\cite{LempickaMirek2024}, providing a way to realise spin-selective directional coupling of lasing states. We further investigate the polarization inheritance properties of the system and realize the unconventional coupling regime between lasing states extending beyond the nearest-neighbour approximation.
Based on the demonstrated properties of our material platform new class of photonic systems can be developed with unexpected functionality of all-optical reconfigurable geometry, in-plane coherence control and electrical tuneability at room temperature, paving the way towards all-optical computing and lattice physics on chip in the weak light-matter coupling regime.

\section{Results}\label{sec2}

\subsection{Platform design and approach}\label{subsec1}

Our lasing device is an electrically tunable planar microcavity operating in the weak light-matter coupling regime containing a LC solution with homogeneously dispersed pyrromethene 580 (P580) laser dye, Fig.~\ref{fig:schematic}(a) [see Methods for more details on the sample and experimental procedure]. Detailed analysis of the light-matter coupling regime in the LCMC is provided in Supplementary Information (SI), Supplementary Note 5. By applying external voltage that changes the spatial orientation the birefringent LC molecular director, we can electrically tune the splitting between the TE and TM cavity modes which can enable intriguing photonic analogues of SOC such as the optical spin-Hall effect~\cite{Lekenta2018, Liang2024} and RD SOC~\cite{Krol2021, empicka-Mirek2022, Muszynski2022}. We define the plane of cavity as the $x$-$y$ plane and the corresponding momentum plane with $k_x$-$k_y$ coordinates. We also define vertical polarization (V) as linear polarization along $y$, and horizontal polarization (H) along $x$. Under applied voltage the LC molecules tilt in the $y$-$z$ plane where $z$ is the out-of-plane direction, or direction of emission.

\begin{figure}[!h]
    \centering
    \includegraphics[width=0.95\linewidth]{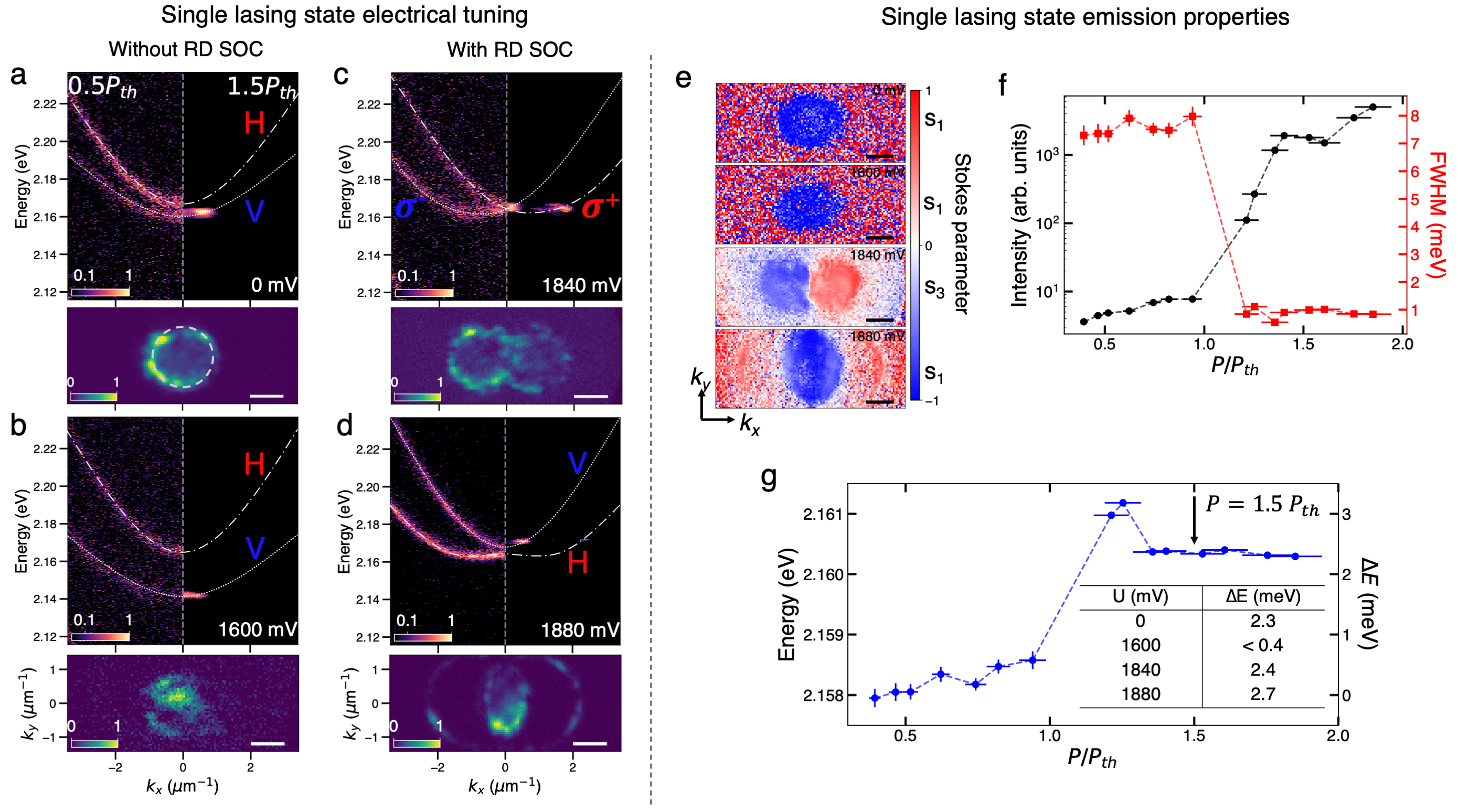}
    \caption{{\bf Single lasing state emission characteristics: electrical tuning and the blueshift-induced in-plane momentum of coherent emission.} {\bf (a-d)} Normalized PL intensity shown in {\bf (above)} energy-resolved momentum-space along $k_y=0$ for four values of external voltage, corresponding to {\bf (a)} 0 mV, {\bf (b)} 1600 mV, {\bf (c)} 1840 mV, and {\bf (d)} 1880 mV, pumped below ($0.5P_{th}$, left) and above ($1.5P_{th}$, right) the lasing threshold for each case and {\bf(below)} momentum-space pumped above ($1.5P_{th}$) the lasing threshold, dot-dashed and dotted white curves show {\bf (a,b,d)} horizontally (H) and vertically (V) polarised or {\bf (c)} right-circularly ($\sigma_+$) and left-circularly ($\sigma_-$) polarised cavity modes; {\bf (e)} momentum-space distribution of the normalised $s_1$ {\bf (first, second and fourth)} and $s_3$ {\bf (third)} Stokes components describing the degree of linear ($s_1$) or circular ($s_3$) polarisation of the emission profile as defined in Sec.\ref{subsec42}, corresponding to momentum-space at {\bf (a-d)}; {\bf (f)} photoluminescence intensity (black circles), full-width at half-maximum (FWHM, red squares), and {\bf (g)} energy and the blueshift ($\Delta E$) of the lasing state versus pump power at zero voltage, extracted from the fitting of the emission spectral profile obtained by integrating dispersion data around $k_x=0$ over $\pm 0.2\mu m^{-1}$ of the energy-resolved momentum space measurements. Error bars of y-axis are extracted from the covariance matrix of the fitting and represent the standard error. Horizontal error bars represent standard deviation of the pump power due to the laser shot-to-shot instability. Table in the inset {\bf (g)} compares the values of the blueshift  at $P=1.5P_{th}$ corresponding to four external voltage values ($U$) in {\bf (a-d)}. Colour scales represent normalized {\bf(a-d)} photoluminescence intensity and {\bf(e)} Stokes parameter (arbitrary units). For better visibility, energy-resolved images {\bf (a-d)} are illustrated in logarithmic scale, saturated below 0.05. Grey dashed ring in momentum-space in {\bf (a)} with the radius ($k_l$) corresponds to the average value of the in-plane photon outflow wavenumber. Rashba-Dresselhaus spin-orbit coupling (RD SOC) regime is present at {\bf (c)} 1840 mV, and {\bf (d)} 1880 mV.}
    \label{fig:singlestate}
\end{figure}

Fig.~\ref{fig:singlestate}(a-d) shows energy-resolved momentum space photoluminescence (PL) of the lasing state below and above the lasing threshold at different voltages, nonresonantly pumped with a single vertically polarized Gaussian spot. Below the threshold (left half of panels) both horizontally and vertically polarised dispersion curves are visible. Despite the linear polarization of the laser pump, the intermolecular energy transfer enables the fast depolarization of the excited states and unpolarized emission below the lasing threshold ~\cite{Chang2007, Musser2017,  Yagafarov_CommPhys2020,Plumhof2014} for the case of relatively small H-V splitting near $k_{\parallel}=0$ comparing to the broadening of the organic dye emission. When external voltage is applied, we observe a shift of the vertically polarised subset of cavity Fabry-P\'{e}rot modes towards higher energy due to the increase of the refractive index coming from the LC reorientation inside the cavity for vertically polarised light, while horizontally polarised modes remain unchanged [compare Fig.~\ref{fig:singlestate}(a) with~\ref{fig:singlestate}(b)]. When two orthogonally linear polarised modes of opposite parity (i.e. different longitudinal mode index) are brought in resonance by electrical tuning, the system enters the RD SOC regime~\cite{Rechcinska_Science2019, Muszynski2022}, shown in Fig.~\ref{fig:singlestate}(c). The dispersion valleys then become circularly polarised with opposite circular polarisations symmetrically shifted in Fourier space along the $k_x$ axis. At sufficiently low momenta, the in-plane dispersion relation for $\sigma^\pm$ circularly polarized photons can be written 
\begin{equation} \label{eq.RDdisp}
E_\pm = \frac{\hbar^2 k_\parallel^2}{2m} \pm \delta_\text{RD} k_x
\end{equation}
where $m$ is the effective photon mass and $\delta_\text{RD}>0$ defines the valley separation $2\delta_\text{RD} m/\hbar^2$. Above the lasing threshold, we observe the collapse of the emission to a narrow energy state laying slightly above the bottom of the cavity photon dispersion curve. The linear polarization of the emission now corresponds to the linear polarisation of the pump as seen in the normalized momentum space $S_1$ Stokes parameters of the emission in Fig.~\ref{fig:singlestate}(e) for the 0 mV and 1600 mV cases [see Eq.~\eqref{eq.Stokes} for definition of the Stokes parameters and connection to the emission polarization]. The transition from unpolarized emission below the lasing threshold to highly polarized emission in the lasing regime is due to the carrier recombination, enabled by the offset of stimulated emission, happening faster than the depolarization effects in the dye. \cite{Yagafarov_CommPhys2020, Plumhof2014}.  In the RD SOC regime at 1840 mV, both circular components are present simultaneously in their respective valleys as seen in the $S_3$ Stokes parameter [see Eq.~\eqref{eq.Stokes} and Fig.~\ref{fig:singlestate}(e)]. Further details on polarisation inheritance are presented in Supplementary Note 1.

Figure~\ref{fig:singlestate}(f,g) presents the dependence of PL intensity, FWHM, and emission energy on the pumping power at 0~mV external voltage. Initially, a linear increase in intensity is observed, followed by a sharp nonlinear behaviour and a step-like drop in FWHM above the threshold pump pulse energy $P_{th}\approx0.6 \mathrm{\mu J}$, both hallmarks of a transition from spontaneous emission to photonic lasing. This transition is further evidenced by a sharp blueshift of the emission relative to the cavity dispersion minimum below threshold. 
In conventional inorganic systems, such as GaAs-quantum well microcavities with Wannier-Mott excitons, such blueshift near threshold is a recognized signature of polariton lasing~\cite{Weihs2004}. However, this interpretation does not straightforwardly apply to organic microcavities with Frenkel excitons, which are highly localized and do not support comparable repulsive interactions. It has been shown~\cite{Yagafarov_CommPhys2020, Putintsev2023} that, in organic systems, the blueshift primarily arises from the saturation of molecular transitions and the associated refractive index change. Hence, it is present in both weak and strong light-matter coupling regimes and, importantly, is wavelength dependent, see Fig~\ref{fig:singlestate}(a-d). Indeed, it has been shown in \cite{Yagafarov_CommPhys2020} that the microcavity modes renormalization upon optical excitation can be described using the general Kramers-Kronig analysis of the organic dye complex refractive index. Assuming the decrease of the imaginary part of the refractive index due to the saturation of optical transitions, the corresponding change of the real part follows in the area of anomalous dispersion. Therefore, the blueshift value depends on the relative energy between the bottom of the cavity photon dispersion band and the resonance energy of the optical transition in the intracavity media, which, in our case, is the absorption of P580 dye in liquid crystal solution \cite{Mowatt2010}. It can be seen in Fig ~\ref{fig:singlestate}(a-d) that the blueshift is present when the vertically polarized cavity mode energy at $k_x=0$ is around 2.16~eV or higher (corresponding to 0 mV, 1840 mV, and 1880 mV case in ~\ref{fig:singlestate}(a,c,d)). Such behaviour is expected assuming the proposed above blueshift mechanism associated with the saturation of the optical transitions, as the P580 dye absorption flattens around 0 below 2.15 eV, while experiencing a strong increase above 2.16 eV \cite{Mowatt2010} (see table in the inset Fig.~\ref{fig:singlestate}(g)), which, upon the saturation of the optical transition, leads to the corresponding decrease of the refractive index and cavity mode renormalization under the optical pumping area. In case of tightly focused ($\approx \ 3 \ \mu m$) optical pumping, it creates sharp gradient of the cavity mode energy, which acts as a localized potential sufficient to induce the outflow of coherent photons with finite average absolute value of in-plane momentum ($k_{||} \neq 0$) radially propagating away from the pumping spot in the cavity plane. It is then evidenced in the ring-shaped profile of emission in momentum-space (or two split rings relative to $k_x=0$ in the RD regime), with the radius $k_{l}$ determined by the amount of pump-induced blueshift [see momentum-space PL at the bottom of Figs.~\ref{fig:singlestate}(a-d)]. We define $k_l$ as an averaged absolute value of the in-plane photon outflow wavenumber, which for the angle-independent emission in $k_x  - k_y$ plane corresponds to the radius of the ring-shaped emission in Fourier space (Fig.~\ref{fig:singlestate}(a)). This ring is similar to a resonant Rayleigh scattering ring when waves are scattered by disorder and couple to many different momenta~\cite{Langbein2002}. Moreover, for tuneable LCMCs, this blueshift depends on the spectral detuning between the P580 dye absorption and the cavity mode, which can be modulated via applied voltage [Fig.~\ref{fig:singlestate}(a,b)]. Consequently, the $k$-space profile of the coherent emission is electrically controllable through the modulation of the local potential. While the underlying mechanisms differ between room-temperature organic microcavities and cryogenically operated GaAs-based systems, both generate coherent emission with finite in-plane momentum, which is a key requirement for mutual phase-locking between lasing sites and the emergence of spatially extended lasing supermodes via in-plane coherent energy exchange.

\subsection{Dyad lasing supermode}\label{subsec_dist}

\begin{figure}[!h]
    \centering
    \includegraphics[width=1\linewidth]{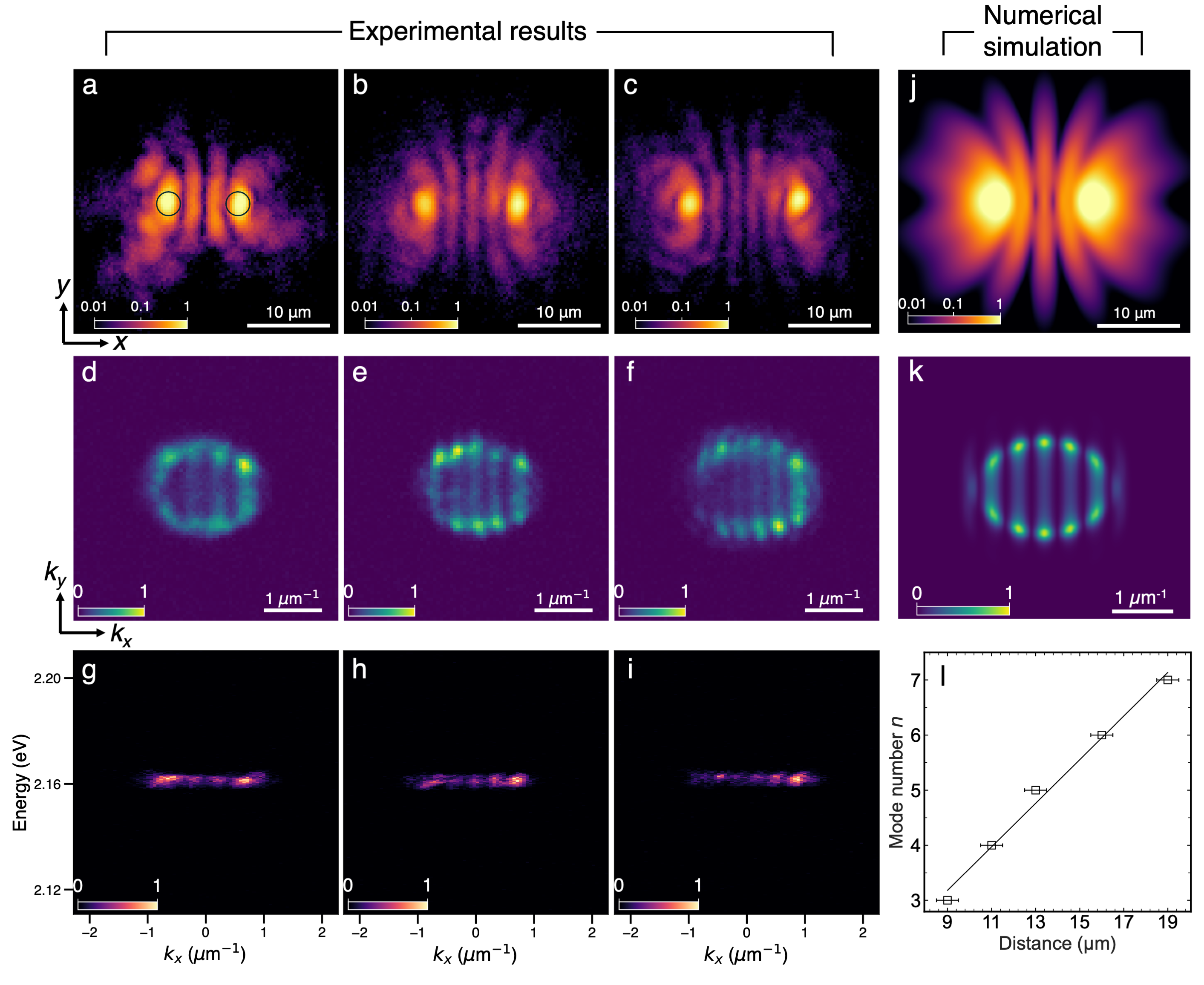}
    \caption{{\bf Dyad supermode lasing state}. {\bf (a-i)} Experimental images of the dyad supermode lasing state emission in {\bf (a-c)} real-space, {\bf (d-f)} momentum space, and energy-resolved momentum-space along $k_y=0$ {\bf (g,h,i)} for two pump spots (black circles) separated by {\bf (a,d,g)} 9 $\mu$m, {\bf (b,e,h)} 11 $\mu$m, and {\bf (c,f,i)} 13 $\mu$m. {\bf (j,k)} Numerically simulated {\bf (j)} real-space and {\bf (k)} momentum-space distribution of the supermode lasing state emission corresponding to the experimental data at {\bf (b,e)}. {\bf (l)} Dependence of the extended lasing state mode number \textit{n} on the pump separation distance. The error bars represent standard deviation of the lasing spot position and are defined by the fragmentation of the lasing state within the pumping spot area. Each spot was pumped with the fixed spot size Gaussian beam at $P_{1,2}=1.5P^{(1)}_{th}$, where $P^{(1)}_{th}$ is the lasing threshold for a single isolated lasing state. Colour scales represent normalized photoluminescence intensity (arbitrary units). For better visibility real-space images {\bf (a-c,j)} are illustrated in logarithmic scale saturated below 0.01.}
    \label{fig:dyad_dist}
\end{figure}

To demonstrate the realisation of coupled extended lasing supermodes formed by phase-locking of two spatially separated lasing modes, we excite the microcavity with vertically polarised two-spot pump profile (i.e., a lasing dyad) with interspot separation distance $d$ in the range of 9 to 19~$\mathrm{\mu m}$, see Fig.~\ref{fig:dyad_dist}.  We keep the microcavity at 0~mV external voltage corresponding to the cavity regime shown in Fig.~\ref{fig:singlestate}(a), where single lasing state occurs with the blueshift $\Delta E$ in the range of 2-3 meV and corresponding in-plane momentum $|k_{l}|\approx 1.0$ $\mathrm{\mu m^{-1}}$ relative to the bottom of the vertical dispersion band $E_0 \approx2.16$~eV. RD SOC is not observed as horizontally and vertically polarised cavity modes are slightly separated in energy.

Fig.~\ref{fig:dyad_dist}(a-f) shows the experimentally measured real- (a-c) and momentum- (d-f) space distributions of the PL from the spatially extended supermode lasing state formed due to the mutual exchange of coherent photons out-flowing from each of the two individually pumped lasing spots and establishing mutual coherence between them via the stimulated emission process.  Once the coupling between the lasing spots is established, they act as coherently coupled oscillators, which relative phase is defined by the distance between them. In the absence of coherent energy exchange, one would observe two independent lasing spots in real space with the PL intensity monotonically decaying outside of the pumping spot areas, Fig.~\ref{fig:schematic}(b), while in Fourier space, one would have a ring-shaped pattern similar to the single state case. Instead, here we observe phase-locking between two individually pumped lasing states, evidenced by a characteristic oscillatory interference pattern of PL intensity in both real and Fourier spaces along the axis connecting the pump spots. This interference pattern directly reveals a macroscopic coherent wavefunction extending over distances up to $d = 20 \mathrm{\mu m}$. The resulting dyad lasing supermodes exhibit either even or odd parity with respect to their reflection axis at $x=0$. The parity is determined by the relative phase between the lasing centres ($\Delta \phi = 0$ or $\pi$), depending on the separation $d$ between the pump spots, as expected and observed previously for polariton dyads~\cite{Topfer2020}. Maximum intensity at the midpoint indicates in-phase synchronisation ($\Delta \phi = 0$), while a minimum of the intensity in the centre reflects anti-phase synchronisation ($\Delta \phi = \pi$). In both cases, a single mode of corresponding parity was present in energy-resolved Fourier-space PL [Fig.~\ref{fig:dyad_dist}(g-i)]. We quantitatively characterise the supermode by introducing the mode number $n$ corresponding to the number of nodes between the pump spots analogous to the mode numbers used to describe the higher-order Hermite-Gaussian transverse modes. A linear increase in the mode number $n$ of the spatially extended supermode is visible when the distance between pumping spots $d$ is increased by approximately half-wavelength of the outflowing photons, that is $k_l d = n \pi$ where $k_l$ is the in-plane photon outflow wavenumber, Fig.~\ref{fig:dyad_dist}(l).

To theoretically reproduce the real and Fourier-space PL from the dyad lasing state, we perform numerical simulations based on a two-dimensional (2D) non-Hermitian Schrödinger equation under drive and dissipation (a simplified version of Maxwell-Bloch equations~\cite{ning1997effective,berloff2013universality}). Details about the theoretical model employed are presented in Methods (Sec.~\ref{subsec43}). In Fig.~\ref{fig:dyad_dist}(j,k) we show the simulation results for the dyad lasing state separated by $d=11\: \mathrm{\mu m}$ corresponding to the experimental data presented in Fig.~\ref{fig:dyad_dist}(b,e). Corresponding simulations of the lasing states for $d= 9\:\mathrm{\mu m}$ and $13\:\mathrm{\mu m}$ can be found in the Supplementary Note 6. The slight mismatch between the experimental and numerically obtained data is due to the local disorder morphology of the sample and is mostly present in real-space PL profiles.

\begin{figure}[!h]
    \centering
    \includegraphics[width=1\linewidth]{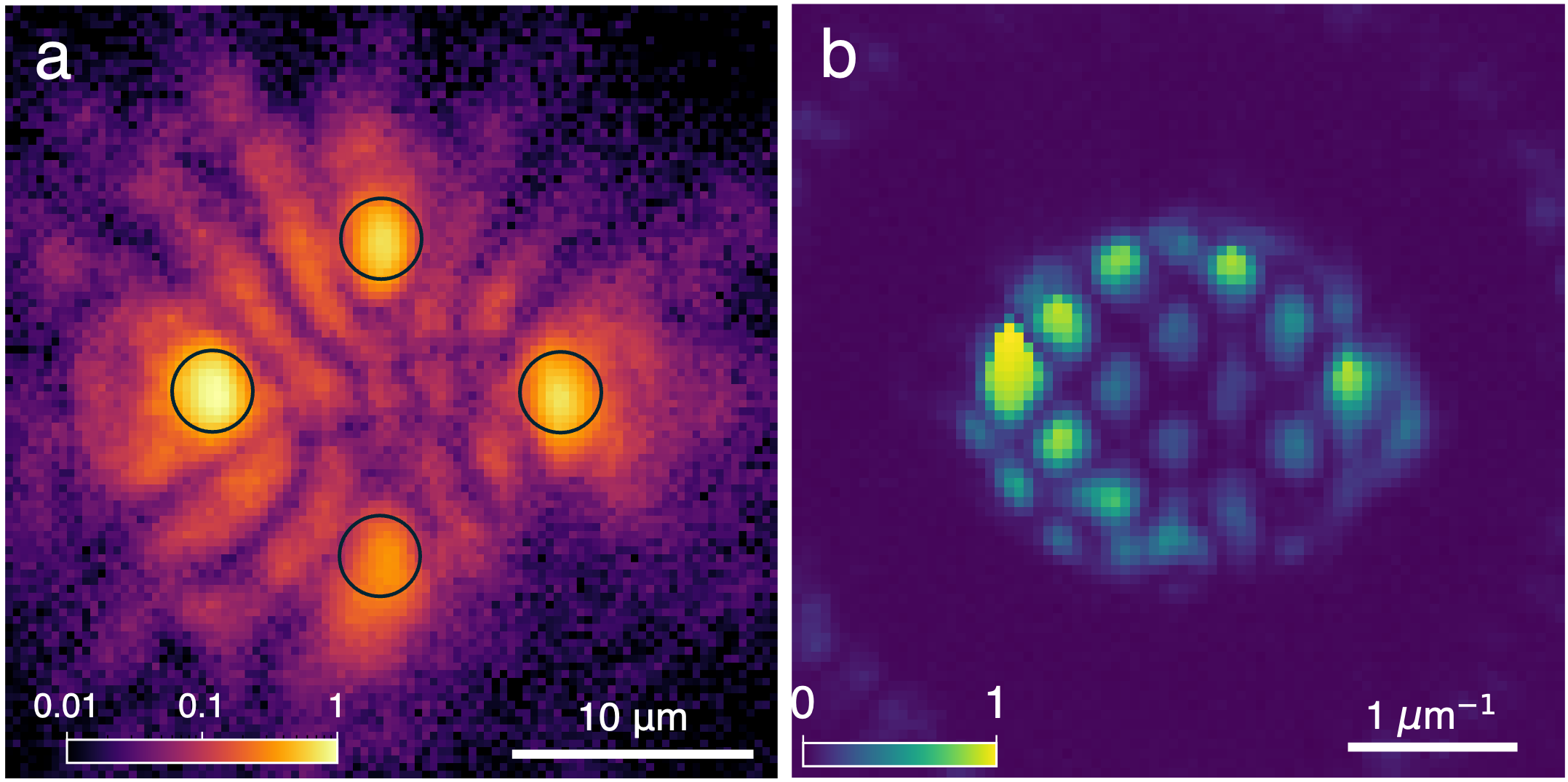}
    \caption{{\bf  Two-dimensional lattice supermode}. {\bf (a-b)} Experimental images of the supermode lasing state emission in {\bf (a)} real-space, {\bf (b)} momentum space for the $2 \times 2$ lattice pump geometry (black circles). Colour scales represent normalized photoluminescence intensity (arbitrary units). 
    }
    \label{fig:4_spots}
\end{figure}

In order to highlight the isotropy of the coupling at 0~V we demonstrate in Fig.~\ref{fig:4_spots} the phase-locking in a square $2 \times 2$ lattice pump geometry (shown here rotated by $45^\circ$ in cavity plane) with anti-phase synchronisation between nearest neighbours ($\Delta \phi_{i,i+1} = \pi$).

The phase-locking between the spatially separated lasing states is further supported by the spatial coherence measurements, see Supplementary Note 2 for details.  We therefore refer to these phase-locked lasing states as a lasing supermode or a dyad supermode in the particular case of two phase-locked states. Such a phase-locking mechanism implies that the resulting mode parity depends not only on the spatial separation but also on the in-plane momentum ($k_{l}$) of the coherent emission. While we have demonstrated the dependence of the supermode parity on distance for a fixed $k_{l}$, the latter can be tuned electrically in our system, as discussed in Sec.~\ref{subsec1}.

\subsection{Electrical control over coupling in a dyad lasing supermode. Coupling in the presence of spin-orbit interaction.}\label{subsec_electrical}

Electrical control of the cavity photon dispersion is one of the major advantages of the LCMCs. As we have shown in Sec.\ref{subsec1}, the blueshift above the lasing threshold is wavelength-dependent and at the specific regimes of cavity detuning it provides a way to alter the value of in-plane momentum ($k_{l}$) of the lasing state emission. Below we demonstrate that the efficiency of coupling in a dyad lasing supermode is defined by this in-plane momentum distribution. Hence, electrical tuneability of the cavity opens a way to reversibly control the coupling between the spatially separated coherent lasing states and to shape the emission of lasing supermode in momentum space.

\begin{figure}[!h]
    \centering
    \includegraphics[width=1\linewidth]{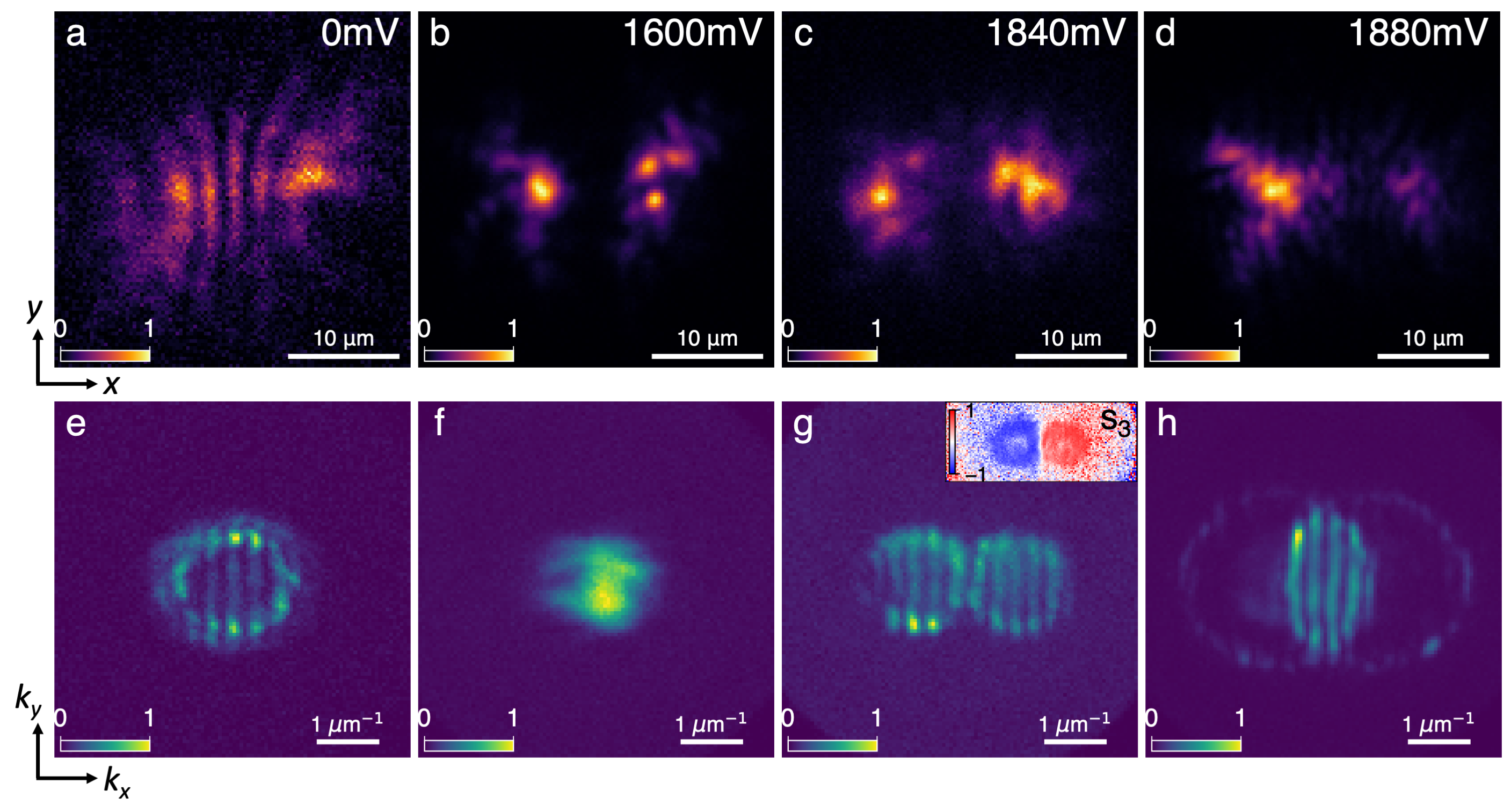}
    \caption{{\bf Electrical control of coupling in a dyad supermode lasing state}. Experimental images of the lasing state emission in {\bf (a-d)} real-space, {\bf (e-h)} momentum space for two pump spots at {\bf (a,e)} 0 mV, {\bf (b,f)} 1600 mV, {\bf (c,g)} 1840 mV and {\bf (d,h)} 1880 mV, showing {\bf (a)} coupled supermode lasing state, {\bf (b)} uncoupled separate lasing spots, and {\bf (c)} and {\bf(d)} extended lasing state in RD regime and slightly detuned from RD regime, respectively. Inset in {\bf (g)} shows the corresponding momentum-space distribution of the $s_3(\boldsymbol{r})$ normalised Stokes component of the emission.  Colour scales represent normalized {\bf(a-h)} photoluminescence intensity and {\bf(inset in g)} $s_3$ Stokes parameter (arbitrary units). Each spot was pumped with the fixed spot size Gaussian beam, separated by the fixed distance of \SI{13}{\micro\meter} at $P_{1,2}=1.5P^{(1)}_{th}$, where $P^{(1)}_{th}$ is the lasing threshold for a single isolated lasing state.} 
    \label{fig:dyad_volt}
\end{figure}

In Fig.~\ref{fig:dyad_volt} we show the experimentally measured real- (a-d) and momentum-space (e-h) distributions of the dyad PL corresponding to four distinctive regimes of coupling between the two pumped lasing spots, controlled by an applied external voltage $V$ ranging from 0 to 1880~mV.

The first regime of interaction is shown in Fig.~\ref{fig:dyad_volt}(a,e) and corresponds to the case of minor energy detuning between horizontally and vertically polarised modes at 0~mV external voltage with the bottom of the vertical dispersion band $E_0 \approx2.16$~eV [Fig.~\ref{fig:singlestate}(a)]. Similar to the dyad supermode described in Sec.~\ref{subsec_dist}, one can observe a bright interference fringe at the mirror symmetry axis ($x=0$ and $k_x=0$) indicating spontaneous phase locking of the lasing nodes into a macroscopic even parity mode. In contrast, when voltage of 1600~mV is applied, see Fig.~\ref{fig:dyad_volt}(b,f), the coupling between the lasing spots is severely reduced, evidenced by a lack of interference fringes in both real and Fourier space. In this scenario, the PL is mostly localized to the pump gain regions in real space with no apparent expanding envelope of outflowing photons, as seen from the corresponding Fourier space PL which displays a wide Gaussian distribution centered around $k_\parallel = 0$ in sharp contrast to the ballistic ring. This is due to the significant negative energy detuning of the vertically polarised cavity mode relative to the zero voltage condition and the corresponding drop of the blueshift ($\Delta E<0.4$~meV) when pump is also vertically polarised. Indeed, a lower pump-induced blueshift for the pumped photons in each lasing state means that the radially expanding part of the wavefunction acquires smaller kinetic energy and therefore slower outflow velocity away from the pump spot. This is seen as a decrease of the in-plane component of the emission $k_l$ in momentum space as discussed in Fig.~\ref{fig:singlestate}(b) for the single pump spot. As a result, the overlap between neighbouring lasing states decreases and their coupling diminishes with consequent lack of interference fringe contrast. 

Interestingly, in Fig.~\ref{fig:dyad_volt}(c,g) we observe revival of the dyad coupling when entering the RD SOC regime at higher voltages $V = 1840$ mV. Here, mixing between two distinct cavity modes results in a dispersion relation made of two off-centred cross-circularly polarized parabolas or valleys, displaced along $k_x$ axis parallel to the axis of the dyad [see Eq.~\eqref{eq.RDdisp} and Fig.\ref{fig:singlestate}(c)]. Since the bottom of each dispersion branch remains similar in energy to that of the vertically polarised cavity mode at $V = 0$~mV, the wavelength-dependent PL blueshift above the threshold is unchanged compared to the initial case [Fig.\ref{fig:singlestate}(a,c)], as is the corresponding ring radius in each valley in Fourier space [Fig.\ref{fig:dyad_volt}(e)]. Here, we observe emission from two separate ballistic rings in Fourier space displaced symmetrically along $k_x$ with opposite circular polarisation corresponding to iso-energy curves of each paraboloid (see the inset in Fig.~\ref{fig:dyad_volt}(g) for $S_3(\mathbf{k}_\parallel)$ Stokes parameter distribution in momentum space). We note that due to RD SOC this regime demonstrates spin-dependent anisotropy of the supermode emission with $\langle k_x^{\sigma+} \rangle > 0 > \langle k_x^{\sigma-} \rangle $ and $\langle k_y^{\sigma+} \rangle =  \langle k_y^{\sigma-}  \rangle = 0$ with a clear fringe pattern observed in Fourier space, proving the phase-locking revival between the two pumped lasing states. Phase-locking revival was also observed in restoring of spatial coherence in the interferometry measurements, see Supplementary Note 3 in SI. In real space the total intensity of the fringes between the pump spots is lowered relative to the case of $V=0$~mV. We attribute this to the real- and Fourier-space redistribution of the phase-locked dyad emission in the presence of spin-orbit coupling when each of the circularly polarized components propagates in opposite directions and has an oscillating intensity pattern of lower intensity, spatially shifted relative to each other. It results in smoothed modulation of the total intensity, while fringe pattern is clearly observed for each circular polarisation separately, see in Supplementary Note 3.

\begin{figure}[!h]
    \centering
    \includegraphics[width=1\linewidth]{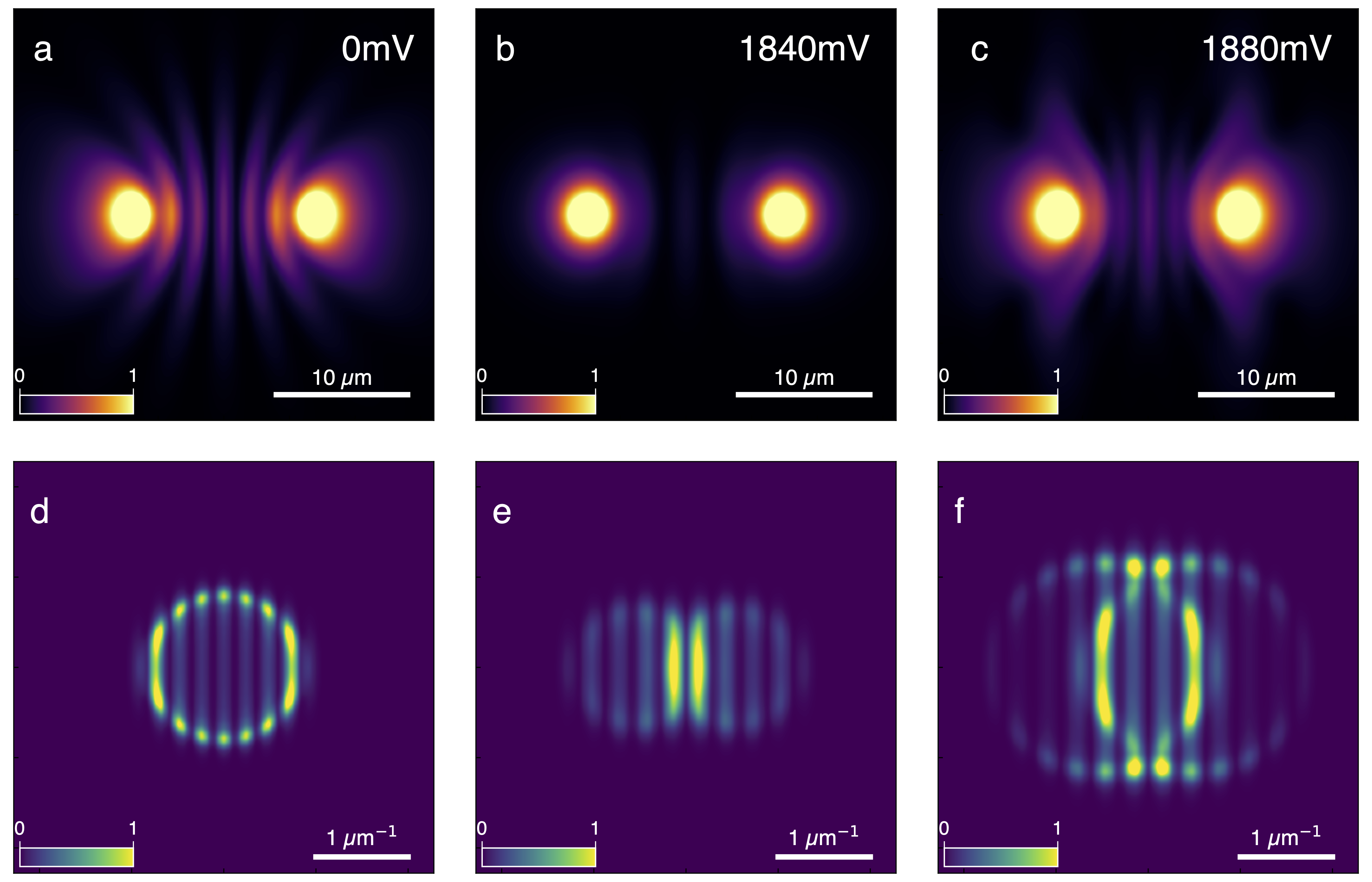}
    \caption{{\bf Simulation for an electrically tuneable coupling in a dyad configuration} corresponding to profiles shown in Fig.~\ref{fig:dyad_volt}, for $V = 0$ $\text{mV}$ \textbf{(a,d)}, $1840$ $\text{mV}$ \textbf{(b,e)} and $1880$$\text{mV}$ \textbf{(c,f)}, respectively. Colour scales represent normalized photoluminescence intensity (arbitrary units).} 
    \label{fig:dyad_voltage_simulation} 
\end{figure}

As the voltage is increased even further to 1880 mV [see Fig.~\ref{fig:dyad_volt}(d,h)], the system is shifted out of the RD SOC regime with a subsequent gap opening at branch crossing point $k_\parallel=0$ with linearly polarized states forming at low momenta. The new dispersion relation, sufficiently close to the RD SOC resonance, can be approximately written as,
\begin{equation}\label{eq.RDdisp2}
E_\pm = \frac{\hbar^2}{2} \left(\frac{k_x^2}{m_x} + \frac{k_y^2}{m_y} \right) \pm \sqrt{(\delta_\text{RD} k_x)^2 + \delta^2} 
\end{equation}
where $m_{x,y}$ are renormalized effective photon masses along the $x$ and $y$ directions~\cite{Rechcinska_Science2019} and $\delta$ describes splitting between horizontal and vertical linear polarized modes at $k_\parallel = 0$. The modified dispersion leads to a higher photon population in the mode with linear polarisation component parallel to the pump polarization. The emission still partially occupies two ballistic circles in the momentum space PL as seen at high wavenumbers in Fig.~\ref{fig:dyad_volt}(h). The radius of the circles is larger since the emission energy is now located further above the minimum of the dispersion~\eqref{eq.RDdisp2}. Interestingly, here the dyad mostly populates the low momentum intersect between the two circles in momentum space corresponding to photons with mostly linear vertical polarization projection. The result is a peculiar distribution of the emission in Fourier space in which the emission anisotropy is reversed relative to the case of pure RD SOC regime ($\langle |k_y| \rangle > \langle |k_x| \rangle$). Control of the in-plane momentum anisotropy is another important feature of our system as it potentially offers a way for the realisation of a direction-dependent coupling in two-dimensional arrays of phase-locked coherent emitters. Similar to the previous case the fringe pattern in real space is distorted and observable mostly around pumping spots. However, it is clearly visible in the Fourier space over the whole area of emission distribution. We verify our interpretation through 2D non-Hermitian Schrödinger (paraxial Maxwell-Bloch) simulations shown in Fig.~\ref{fig:dyad_voltage_simulation}, see Sec.~\ref{subsec43} for details. We note that the experimentally observed effect of reduced interference fringes visibility in the RD SOC regime, accompanied by the clear phase-locking in Fourier space, is similarly observed in numerical simulations.

\subsection{Unconventional coupling in 1D lasing supermode}\label{subsec_NNN}

\begin{figure}[!h]
    \centering
    \includegraphics[width=1\linewidth]{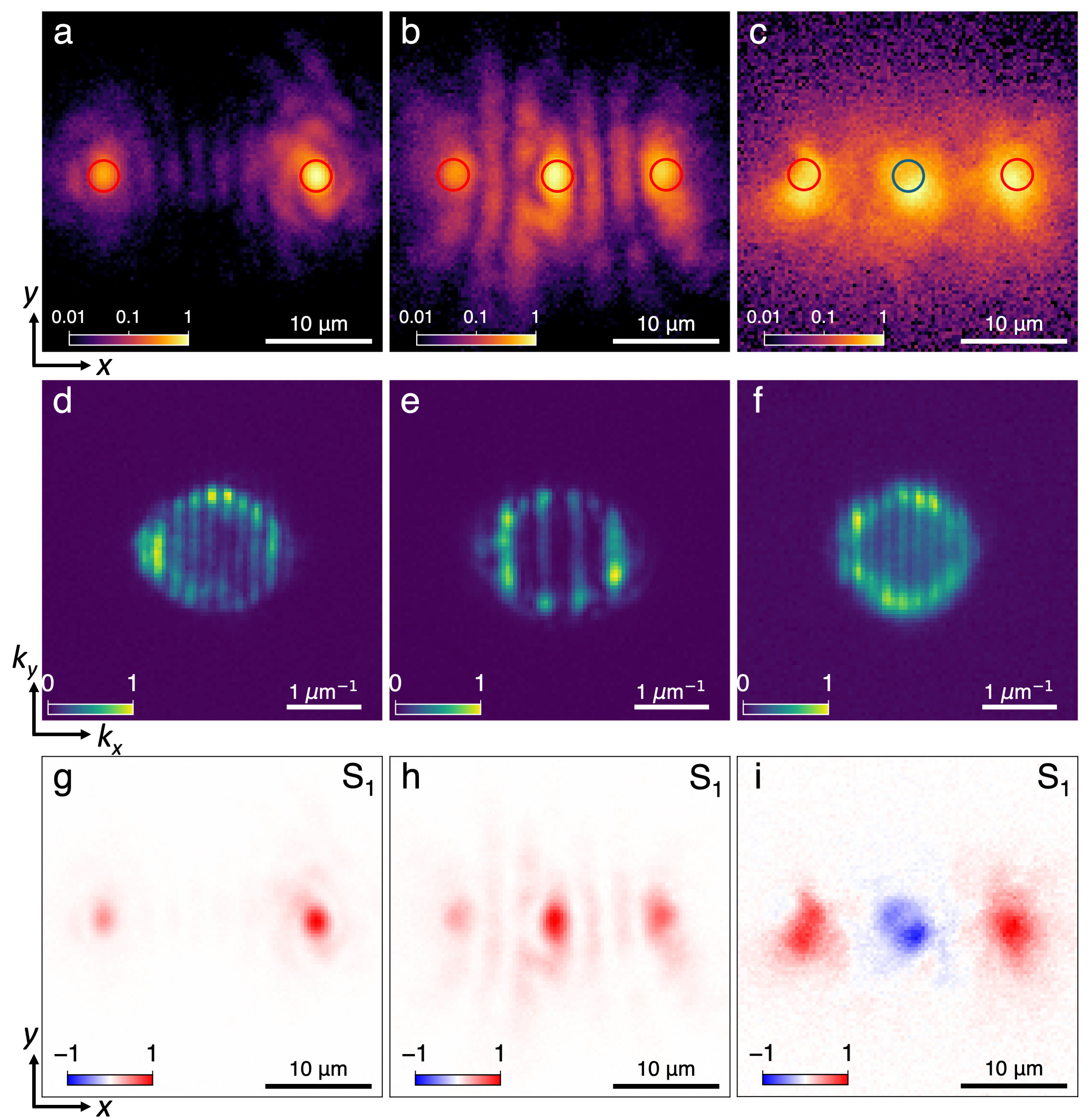}
    \caption{{\bf Unconventional coupling in a 1D chain of pump spots}. Experimentally measured {\bf (a-c)} real-space, {\bf (d-f)} momentum space, and {\bf (g-i)} real-space non-normalized $S_1$ component of the Stokes vector of the lasing state emission. {\bf (a,d,g)} Left panel shows data corresponding to the two pump spots at 20 $\mu$m separation distance with horizontally polarized pump (red circles), {\bf (b,e,h)} central panel shows data for three horizontally polarized pump spots separated by 10 $\mu m$, {\bf (c,f,i)} right panel shows data for three pump spots separated by 10 $\mu m$ with central spot pumped with vertically polarized light (blue circle). All measurements were done at zero external voltage. Colour scales represent normalized {\bf(a-f)} photoluminescence intensity and {\bf(g-i)} $S_1$ Stokes parameter (arbitrary units). For better visibility real-space images {\bf (a-c)} are illustrated in logarithmic scale saturated below 0.01.
    } 
    \label{fig:triad}
\end{figure}

In coherently coupled photonic systems the nearest-neighbour (NN) coupling dominates over the next-nearest-neighbour (NNN), including systems based on coupled lasers~\cite{Pal2020}, VCSELs arrays, polaritonic and photonic lattices~\cite{Parto2020}. This is easily understandable, as the coupling strength is usually defined by the mutual overlap of the coupled states wavefunctions, which decay exponentially with the distance from each state centre. Overcoming this behaviour is of prior importance for universal applications of coherently coupled photonic arrays as all-optical simulators ~\cite{Kalinin2022}. The possibility of NNN coupling with polaritonic condensates in GaAs-based microcavities at cryogenic temperatures has been shown previously using distance-dependant spin-screening of NN interaction leveraging the optical spin Hall effect~\cite{Dovzhenko2023}. Same approach based on the TE-TM field is not straightforwardly applicable in organic microcavities, but it can be adapted, because the polarisation basis in these systems is defined by horizontal and vertical components, following the orientation basis of the molecular dipole moment. As we show above, the polarisation of the lasing state is then inherited from the pump polarisation and is linear as long as SOC remains small, see Sec.~\ref{subsec1}. As a result, the efficient coupling between the lasing spots is only possible while they share the same linear polarisation inherited from the pumps. In Fig.~\ref{fig:triad} we demonstrate the realisation of NNN coupling regime in a chain of 3 coupled lasing states, by measuring the real space PL (a-c), momentum space PL (d-f), and (g-i) real space $S_1$ Stokes parameter of the non-normalized Stokes vector of the extended lasing state emission [see Eq.~\eqref{eq.Stokes}]. The corresponding numerical simulations obtained via the 2D non-Hermitian Schrödinger model (Sec.~\ref{subsec43}) are shown in the Supplementary Note 6.

 First, we project the two-spot excitation pattern similar to the one shown in Sec~\ref{subsec_dist} with two horizontally polarised pumping spots separated by $d=20$ $\mu$m distance, see Fig.~\ref{fig:triad}(a,d,g). The extended horizontally polarised lasing supermode of the odd parity ($\Delta\phi =\pi$ phase difference) is formed as evidenced in both real- and Fourier-space. We then project an additional horizontally polarised pumping spot in the middle between the initial two spots, making pump spots pairwise separated by $d/2=10$ $\mu$m, as shown in Fig.~\ref{fig:triad}(b,e,h). The triad supermode coherent state is verified in both real and Fourier space with a $\pi$ phase difference between each pair of the phase-locked lasing states. Interestingly, two subsets of fringes are distinguishable in the Fourier-space. One with the same period of the fringes as for a dyad but with a lower relative intensity, showing phase-locking between outer states and the second at a higher intensity and a period twice as large, showing strong phase-locking between each pair of the nearest neighbours.

However, when we flip the polarisation of the central pump spot to the vertical one, both real and Fourier-space distributions change dramatically, see Fig.~\ref{fig:triad}(c,f,i). First, the fringes in the real space are not visible, due to the low intensity similar to the case of the far-separated dyad [Fig.~\ref{fig:triad}(a)]. The Fourier space shows a superposition of two states, see Supplementary Note 4 for details. The first originates from a NNN type of coupling between the horizontally polarised edge states, which is visible with the odd parity and the same periodicity as the one observed for the far-separated dyad. The second mode is ring-shaped and corresponds to the Fourier space profile of an isolated vertically polarised central lasing state. Thus we eliminate the coupling between the nearest neighbours by exciting the lasing state in the middle with the opposite polarisation to the outer pump spots, while the NNN coupling is still present.

\section{Discussion}\label{sec12}

In conclusion, we have demonstrated the emergence of spatially extended lasing supermode states in a LC dye-filled microcavity with all-optical control over coherent emission in both near- and far-field. We achieve the phase-locking between spatially separated independently pumped lasing states by realising mutual coherent photon exchange leveraging the excitation of high in-plane momentum lasing states at each pumping spot. The latter is realised by embedding within the microcavity a medium capable of exhibiting an optically induced blueshift localised at the excitation spot, which causes the in-plane propagation of coherent photons from each lasing state. By tuning the separation distance between the lasing states, we demonstrate phase-locking with either 0 or $\pi$ phase difference extending over distances up to $d = 20 \mu\mathrm{m}$. Moreover, by applying an external voltage and achieving effective wave retardation in an anisotropic LCMC, we have introduced electrical control over coupling strength between individually pumped lasing states. Thus we alter the intrinsic properties of the supermode extended lasing state by shaping its near- and far-field distributions.

We have also shown the switching on and off mechanism of coupling into spatially extended lasing supermode in a reversible manner, along with the ability to bring anisotropy in the Fourier space distribution of the PL in both $k_x$ and $k_y$ directions. We note that previously reported electrical tuning capabilities in similar LCMC systems \cite{LempickaMirek2024, Muszynski2022} concerned the emission properties of an individual lasing state or polariton condensate. The wavelength tuning arose from the cavity-mode renormalization induced by the rotation of highly-birefringent LCs under an external electric field, while the polarization and condensation threshold tuning originated from the realization of the RD SOC regime. In contrast, in the present study we present the electrical tunability of the in-plane phase-locking mechanism between multiple lasing spots. While the system remains above the lasing threshold in all operation regimes, the in-plane phase-locking between the separately pumped lasing spots is controlled electrically by modifying the Fourier-plane emission distribution of each lasing state. This unique functionality is enabled by the combined effect of electrically tunable cavity modes and the wavelength-dependent blueshift of the coherent emission forming a highly localized optically-induced potential. Moreover, in the RD SOC regime, the emission from the supermode lasing state becomes circularly polarised, the coupling becomes spin-selective and the supermode emission with opposite spin separates in Fourier space. These findings open prospects to create fully optical systems with spin propagation control employing external voltage and optical excitation profile engineering. Utilising the inheritance of the pump polarisation by the supermode coherent emission, we experimentally demonstrate a non-trivial regime of coupling in a chain of phase-locked extended lasing states. In particular, we show that next-nearest-neighbour coupling can be tuned stronger than the nearest-neighbour coupling, which is an important limitation in the majority of experimental platforms for the investigation of many-body phenomena.  With this we establish organic liquid-crystal microcavity as a new versatile electrically tuneable material platform for optical lattice physics at room temperature.  We have focused here on the properties of small one- and two-dimensional supermodes consisting of up to four lasing states. However, in the future larger two-dimensional lattices of coupled lasing states can be developed with the prospect of exploring non-Hermitian physics, topologically protected states \cite{Tsesses2018, Yang2022}, and as a robust platform for photonic simulator \cite{Mahler2024, Parto2020, Tao2022} and all-optical neural networks \cite{Zuo2019, Chen2023, Matuszewski_PhysRevApplied.21.014028_2024} operating in a weakly coupled regime at room temperature and offering the benefits of electrical tuning of the wavelength, polarization, and coherence as well as non-trivial coupling capabilities between the lattice nodes.

\section{Methods}\label{sec4}

\subsection{Fabrication of the dye-filled microcavity with LC molecules}\label{subsec41}

The microcavity consists of two Distributed Bragg Reflectors (DBRs) with the solution of highly birefringent nematic liquid crystal and P580 dye molecules sealed in the cavity region between them. Each DBR is
composed of six pairs of SiO$_2$/TiO$_2$ layers, centered at $\lambda =
560$~nm (2.21~eV), and deposited on a 26~nm transparent ITO electrode on
a glass substrate to allow for electrical control over the orientation of LC molecular director by applying an external voltage, see schematic of the device in Fig.~\ref{fig:schematic}(a). The DBRs are assembled using adhesive containing silica spacers, ensuring a cavity thickness varying wedge-like between 2.5~$\mu$m and 3~$\mu$m. Antiparallel orienting layers (SE-130, Nissan Chem., Japan) were applied to both substrates via spin-coating. The cavity was then filled by capillary action with a high-birefringence, home-made liquid crystal mixture, LC2091* (refractive indices: $n_o = 1.57$, $n_e = 1.98$ at 589~nm, sodium D-line) in the nematic phase, with pyrromethene 580 (P580) laser dye (1\% by wt.) dispersed in the mixture. The elongated shape of the LC molecules provides refractive index anisotropy in {\it x-y} plane of the cavity $(\Delta n(V=0)=n_e-n_o=0.41)$, while external voltage applied to the ITO contact rotates the LC director in {\it y-z} plane of the cavity~\cite{Dabrowski2013, Miszczyk2018}. Consequently, control over the energy of the vertical linearly polarised set of optical cavity modes parallel to the {\it y}-axis is achieved~\cite{Lekenta2018}. The set of horizontal linearly polarised modes parallel to the {\it x}-axis remains unchanged [Fig.~\ref{fig:singlestate}(a-d)].

\subsection{Optical spectroscopy measurements}\label{subsec42}
All measurements were performed at room temperature. We used a non-resonant linearly polarized Q-switched laser excitation at 532 nm wavelength with 5 ns pulse length (Opolette SE 355 LD). All photoluminescence measurements were acquired in a transmission geometry with a 50 mm focal length lens used for excitation. A spatial light modulator was used to shape the spatial profile of the pump beam into the array of Gaussian spots with full-width-at-half-maximum (FWHM) of $\approx3 \ \mu$m with the possibility of adding another pump spot provided by a parallel optical path with polarisation controlled independently by the $\lambda/2$ waveplate.

Output emission was collected using a 50x Mitutoyo microscope objective with a numerical aperture of 0.42 and coupled to a 550 mm spectrometer (Horiba Triax 550) equipped with a 300 grooves mm$^{-1}$ grating and a mirror to provide spectral, wave-vector or spatial resolution depending on the set of lens used to focus the output on the entrance slit of the spectrometer and the width of the slit, with spectral resolution limit of $\sim 0.37 $ meV/pixel. All measurements above the threshold were carried out in a single-shot regime, while below the threshold we accumulated the signal for over 100 pulses.

The non-normalized Stokes parameters of the cavity emission are extracted using the following equation:

\begin{eqnarray}\label{eq.Stokes}
S_{1,2,3}({\bf r})=I_{\mathrm{H, D},\sigma^+}({\bf r})-I_{\mathrm{V, A},\sigma^-}({\bf r}),
\end{eqnarray}
where $\mathbf{r} = (x,y)$ is the in-plane coordinate and  $I_{\mathrm{H(V),D(A)},\sigma^+(\sigma^-)}({\bf r})$ corresponds to horizontally (vertically), diagonally (antidiagonally), and right-circularly(left-circularly) polarized (RCP and LCP for short) PL, respectively. The $S_1({\bf r})$, $S_2({\bf r})$, and $S_3({\bf r})$ components represent the degree of linear, diagonal, and circular polarisations, correspondingly. The total emission is written $S_{0}({\bf r})=I_{\mathrm{H, D},\sigma^+}({\bf r})+I_{\mathrm{V, A},\sigma^-}({\bf r})$ and corresponding normalized Stokes parameters defined as $s_{1,2,3}({\bf r}) = S_{1,2,3}({\bf r}) / S_0$.

\subsection{Theoretical analysis: Non-Hermitian 2D Schrödinger Model (paraxial Maxwell-Bloch theory)}\label{subsec43}

To describe the dynamics and coupling between spatially separated lasing states in the dye-filled LC microcavity [Fig.~\ref{fig:schematic}(a,b)], we propose a set of non-Hermitian 2D Schrödinger equations (NHSE) representing a paraxial limit of Maxwell-Bloch equations~\cite{ning1997effective,berloff2013universality} with additional terms based on rate equations describing a BODIPY-G1 dye-filled microcavity~\cite{Yagafarov_CommPhys2020}, which shares structural similarities with the P580 laser dye employed here. Under the slowly varying envelope approximation, we consider the photonic field within the LC microcavity in the linear polarization basis as $\bm{\psi} = (\psi_\parallel,\psi_\perp)^\text{T}$, where $\psi_\parallel$ and $\psi_\perp$ correspond to photons linearly polarized parallel and perpendicular to the nonresonant pump spots, respectively.

Without any loss of generality, we define the $x$ and $y$ directions in the cavity plane as our polarization basis and refer to them as the ``horizontal'' $\psi_\mathrm{H}$ and ``vertical'' $\psi_\mathrm{V}$ components. This choice yields the following system of equations:
\begin{eqnarray} \label{eq.OSHE}
    i \hbar \frac{\partial \psi_{\mathrm{H,V}}}{\partial t} & = & -\frac{\hbar^2}{2} \left\{\left[ \partial^2_x \left(\frac{1}{m} \pm \frac{1}{\mu} \right)  + \partial^2_y \left( \frac{1}{m} \mp \frac{1}{\mu}\right) \right]   + \hbar G (P_{\mathrm{H}} + P_{\mathrm{V}})\right.\nonumber\\ 
    & + & \left.\frac{i \hbar}{2} (\gamma_{\textrm{xp}} n_{\textrm{H,V}} -\gamma_p)\right\} \psi_{\mathrm{H,V}} 
    \pm  i \frac{\hbar^2}{\mu} \partial_x \partial_y \psi_{\textrm{V,H}} \pm \delta \psi_{\textrm{H,V}}.
\end{eqnarray}
Here, $m$ denotes the effective photon mass within the microcavity, and $\mu$ represents the mass term accounting for the $x$-$y$ polarization anisotropy induced by the transverse electric and magnetic (TE-TM) splitting of the birefringent LC medium~\cite{Rechcinska_Science2019}. This birefringence also causes an energy splitting $\delta$ between the horizontally and vertically polarized modes. Moreover, $\gamma_{\mathrm{xp}}$ is the decay rate describing the transfer of excitons from the non-resonantly pumped reservoir into the microcavity lasing mode, and $\gamma_p$ characterizes the radiative losses from the lasing state to the outside of the microcavity. 

The dynamics of the excitonic reservoir with density $n_{\mathrm{H,V}}$ locally excited by a non-resonant Gaussian pump $P_{\mathrm{H,V}} = p_{\mathrm{H,V}} e^{-r^2/2w^2}$, with amplitude $p_{\mathrm{H,V}}$ and width $w$, is accounted for by the following rate equation:
\begin{equation}
            \frac{\partial n_{\mathrm{H,V}}}{\partial t} = P_{\mathrm{H,V}}  -  (2\gamma_{\mathrm{XX}} + |\psi_{\mathrm{H,V}}|^2 \gamma_{\mathrm{xp}})n_{\mathrm{H,V}}  +   \gamma_{\mathrm{XX}} n_{\mathrm{V,H}}, \label{eq:reservoir}
\end{equation}
The parameter $\gamma_{\mathrm{XX}}$ describes the intermolecular energy transfer rate between $n_{\mathrm{H}}$ and $n_{\mathrm{V}}$, characteristic of organic laser dyes~\cite{Yagafarov_CommPhys2020}. The nonradiative relaxation of the excitonic reservoir $\gamma_\text{NR}$ is safely neglected due to its smaller size compared to other decay rates within the model.

For the dyad lasing states under finite external voltage, the microcavity modes are within the Rashba-Dresselhaus SOC regime~\cite{Rechcinska_Science2019}. In this case, the polarization anisotropy is dominated by a term distinct from the TE-TM splitting as in Eq.~\eqref{eq.OSHE}, and the dynamics of lasing states is described by the following equation:
\begin{equation} \label{eq.RD}
    i \hbar \frac{\partial \psi_{\mathrm{H,V}}}{\partial t} = \left[ -\frac{\hbar^2 \nabla^2}{2m} + \hbar G (P_{\mathrm{H}} + P_{\mathrm{V}})  \pm \delta +\frac{i \hbar}{2} (\gamma_{\mathrm{xp}} n_{\mathrm{H,V}} -\gamma_p) \right] \psi_{\mathrm{H,V}} \pm i \delta_\text{RD} \partial_x \psi_{\mathrm{V,H}},
\end{equation}
with $\nabla^2 \equiv \partial_x^2 + \partial_y^2$, and  $\delta_\text{RD}$ being the strength of the Rashba-Dresselhaus splitting controlled by the applied external voltage.

In theoretical descriptions of typical inorganic semiconductor microcavities, corresponding NHSE commonly depict defocusing Kerr-like $\chi^3$ nonlinear terms for characterizing energy blueshifts~\cite{Wouters_PRL2007}. These explicit nonlinearities can be neglected here, as organic-filled microcavities are described in terms of highly localised Frenkel excitons, reducing pairwise interactions~\cite{Daskalakis_NatMat2014,Yagafarov_CommPhys2020}. However, other nonlinear sources responsible for pump-induced blueshifts are present in organic dye-filled microcavities, such as saturation of optical transitions~\cite{Yagafarov_CommPhys2020}. In the case of LC microcavities, the blueshift can also result from the microcavity mode energy renormalization due to changes in the intracavity refractive index. To address these blueshift mechanisms, we introduce the phenomenological parameter $G$ in Eqs.~\eqref{eq.OSHE} and \eqref{eq.RD}, whose value is set to fit the experimental measurements.

Based on the NHSE, we numerically obtain real-space and momentum space emission profiles for lasing states, considering both dyad and triad lasing modes (Secs.~\ref{subsec_dist} and \ref{subsec_NNN}), as well as the case in which the coupling between the dyad lasing supermode is electrically tuned via external voltage (Sec.~\ref{subsec_electrical}). All emission profiles were obtained considering external non-resonant pump amplitudes strong enough to ensure convergence to a finite-valued state $|\psi_{H,V}|\neq0$ in the steady-state regime $\partial_t |\psi_{H,V}| = 0$. Initial conditions were chosen to be white-noise with a random seed based on the time of day.

For the case of dyad lasing supermodes shown in Fig.~\ref{fig:dyad_dist} corresponding to zero-voltage, we extracted the following parameters: microcavity photon effective masses $m$ = 0.0463 meV ps$^{2}$ $\mu$m$^{-2}$ and $\mu$ = 0.1835 meV ps$^{2}$ $\mu$m$^{-2}$, exciton reservoir to microcavity lasing mode decay rate $\gamma_p$ = 5.2632 ps$^{-1}$, lasing state radiative losses $\gamma_{\mathrm{xp}}$ = 0.02 $\mu$m$^2$ ps$^{-1}$ and TE-TM energy splitting $\delta$ = 4.5 meV. We consider the intermolecular energy transfer ratio $\gamma_{\mathrm{XX}}$ = 0.033 ps$^{-1}$ corresponding to a BODIPY-G1 dye-filled microcavity~\cite{Yagafarov_CommPhys2020}, which shares a similar structure with the P580 laser dye employed here. The results from our numerical simulation for both real- and momentum-space emission are shown in panels (j) and (k) of Fig.~\ref{fig:dyad_dist}, respectively.

For the situation of electric control of coupling in the dyad configuration depicted in Fig.~\ref{fig:dyad_volt}, we used the same parameters above for obtaining the corresponding simulations shown in Fig.~\ref{fig:dyad_voltage_simulation}, except for the effective mass $m$ = 0.048 meV ps$^{2}$ $\mu$m$^{-2}$, $\delta = 0$ and the strength of Rashba-Dresselhaus splitting $\delta_\text{RD}$ = 7 meV $\mu$m for $V = 1840$ mV. The same value of $\delta_\text{RD}$ is extracted for $V = 1880$ mV, but with an additional horizontal-vertical linear polarization splitting at $\delta = \SI{2.0}{\milli \electronvolt}$. In all simulations, we set the phenomenological parameter $G = \SI{0.1}{\micro \meter^2}$ to mimic nonlinear energy blueshifts.

\section*{Data availability}

The data that support the findings of this study are available at Pure research information system (https://doi.org/10.5258/SOTON/D3907).

\bmhead{Supplementary information}

Supplementary information provides additional details on voltage, and pump polarisation dependence of a single lasing state emission, interference measurements, and additional data on unconventional coupling. It also provides additional numerical simulations of the NHSE, and estimation of light-matter coupling strength within our system.  

\bibliography{bibliography}

\section*{Acknowledgements}

The authors acknowledge the support of the European Union’s Horizon 2020 program through a FET Open research and innovation action under grant agreement no. 964770 (TopoLight). D.D. acknowledges support from UK Research and Innovation (Future Leaders Fellowship, Grant Reference MR/V023845/1). L.S.R. acknowledges the Icelandic Research Fund (Rannís), grant No. 239552-051. J.Sz. and L.S.R. acknowledge the project No 2023/51/B/ST3/03025 by the Polish National Science Centre. H.S. acknowledges the projects No. 2024/55/B/ST3/02954 and 2022/45/P/ST3/00467 co-funded by the Polish National Science Centre and the European Union Framework Programme for Research and Innovation Horizon 2020 under the Marie Skłodowska-Curie grant agreement No. 945339.
K.S. and S.D.L. acknowledge the Leverhulme Trust, grant No. RPG-2022-037. P.N. acknowledges the grant No 2024/53/B/ST11/04193 by the Polish National Science Centre.

\section*{Author Contributions Statement}

D.D. conceived the project, designed the experiments, and implemented the experimental setup;
D.D. and K.S. conducted the single-shot measurements;
D.D. and P.Ko. analysed the experimental results;
P.Ko., D.S., G.M. contributed with the results interpretation and data analysis, developed approach to determine the type of light-matter coupling regime;
L.S.R. and H.S. performed theoretical modelling;
P.N., P.M., W.P., P.Ku. designed and fabricated the liquid-crystal dye-filled microcavity and supervised sample fabrication within the project;
M.M., P.Ka. characterised the sample after fabrication;
D.D., L.S.R., H.S wrote the paper with input from all authors.
D.D., J.S., and S.D.L. supervised the research.

\subsection*{Competing Interests Statement}
The authors declare no competing interests.

\end{document}


\title[Article Title]{Supplemental information: Electrically Reconfigurable Extended Lasing State in an Organic Liquid-Crystal Microcavity}


\author*[1]{\fnm{Dmitriy} \sur{Dovzhenko}}\email{dovzhenkods@gmail.com}
\author[2]{\fnm{Luciano} \sur{Siliano Ricco}}
\author[1]{\fnm{Krzysztof} \sur{Sawicki}}
\author[3]{\fnm{Marcin} \sur{Muszy\'nski}}
\author[4]{\fnm{Pavel} \sur{Kokhanchik}}
\author[3]{\fnm{Piotr} \sur{Kapu\'sci\'nski}}
\author[5]{\fnm{Przemys\l{}aw} \sur{Morawiak}}
\author[5]{\fnm{Wiktor} \sur{Piecek}}
\author[6]{\fnm{Piotr} \sur{Nyga}}
\author[5]{\fnm{Przemys\l{}aw} \sur{Kula}}
\author[4,8]{\fnm{Dmitry} \sur{Solnyshkov}}
\author[4]{\fnm{Guillaume} \sur{Malpuech}}
\author[2,3]{\fnm{Helgi} \sur{Sigurðsson}}
\author[3]{\fnm{Jacek} \sur{Szczytko}}
\author[1,9]{\fnm{Simone} \sur{De Liberato}}

\affil[1]{\orgdiv{School of Physics and Astronomy}, \orgname{University of Southampton}, \orgaddress{\street{University Road}, \city{Southampton}, \postcode{SO17 1BJ}, \country{United Kingdom}}}

\affil[2]{\orgdiv{Science Institute}, \orgname{University of Iceland}, \orgaddress{\street{Dunhagi-3}, \city{Reykjavik}, \postcode{IS-107}, \country{Iceland}}}

\affil[3]{\orgdiv{Institute of Experimental Physics, Faculty of Physics}, \orgname{University of Warsaw}, \orgaddress{\street{ulica Pasteura 5}, \city{Warsaw}, \postcode{PL-02-093}, \country{Poland}}}

\affil[4]{\orgdiv{Institut Pascal}, \orgname{Universit\'e Clermont Auvergne, CNRS}, \orgaddress{\street{ClermontINP}, \city{Clermont-Ferrand}, \postcode{F-63000}, \country{France}}}

\affil[5]{\orgdiv{Institute of Applied Physics}, \orgname{Military University of Technology}, \orgaddress{\street{S. Kaliskiego 2}, \city{Warsaw}, \postcode{00-908}, \country{Poland}}}

\affil[6]{\orgdiv{Institute of Optoelectronics}, \orgname{Military University of Technology}, \orgaddress{\street{S. Kaliskiego 2}, \city{Warsaw}, \postcode{00-908}, \country{Poland}}}

\affil[7]{\orgdiv{Institute of Chemistry}, \orgname{Military University of Technology}, \orgaddress{\street{S. Kaliskiego 2}, \city{Warsaw}, \postcode{00-908}, \country{Poland}}}

\affil[8]{\orgdiv{Institut Universitaire de France},  \orgaddress{\city{Paris}, \postcode{F-75231}, \country{France}}}

\affil[9]{\orgdiv{Istituto di Fotonica e Nanotecnologie}, \orgname{Consiglio Nazionale delle Ricerche (CNR)}, \orgaddress{\street{Piazza Leonardo da Vinci 32}, \city{Milano}, \postcode{20133}, \country{Italy}}}

\maketitle

\section{Details on the sample properties and experimental results}

\subsection{Individual lasing state emission properties}

\begin{figure}[!h]
    \centering
    \includegraphics[width=1\linewidth]{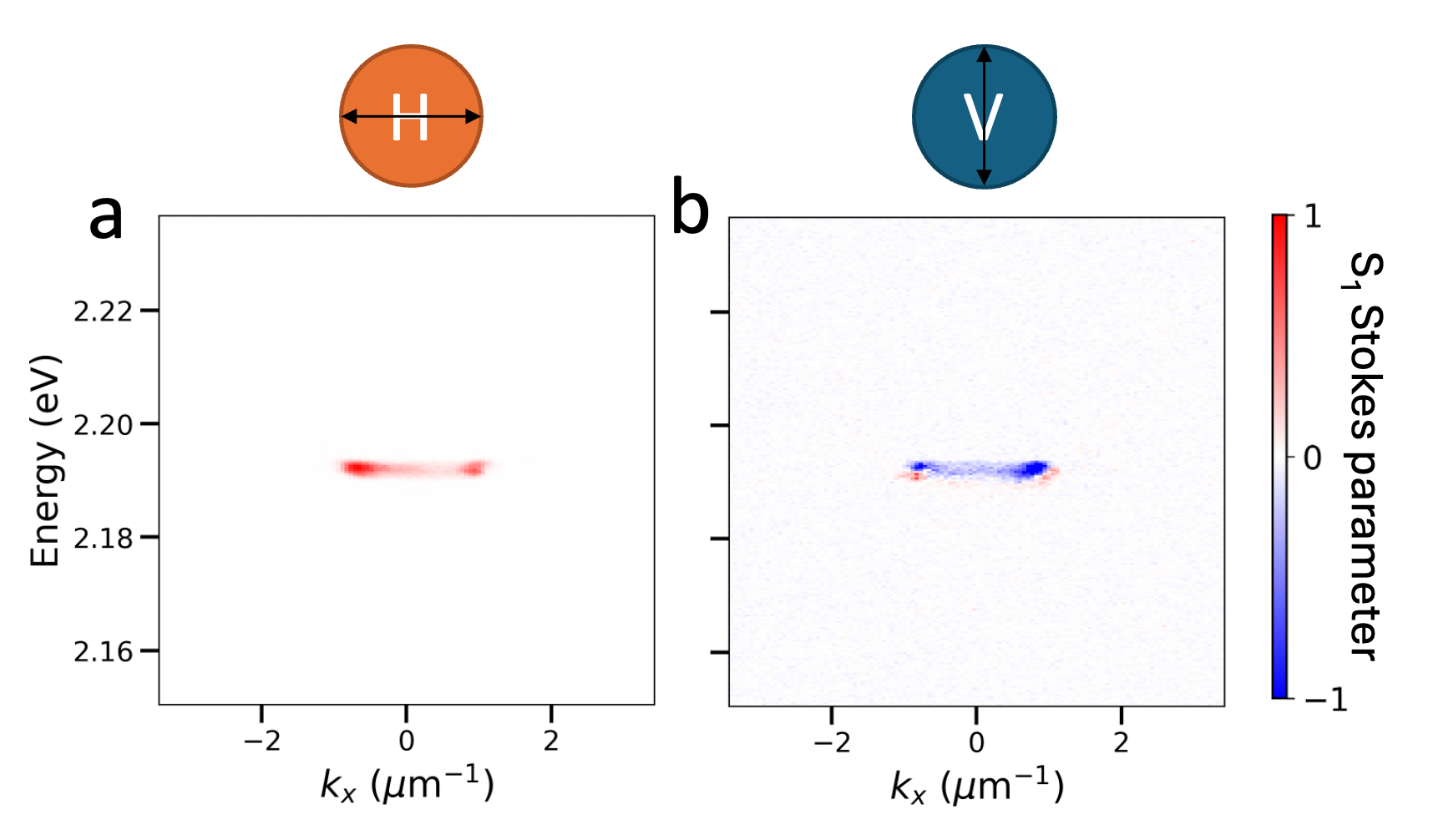}
    \caption{{\bf Inheritance of the lasing state polarisation}. {\bf (a,b)} Energy-resolved along $k_y=0$ momentum-space distribution of the non-normalised $S_1$ Stokes component of the emission from a single spot pumped with {\bf (a)} horisontally and {\bf (b)} vertically polarised light. Schematic above shows the respective polarisation of the pump}
    \label{fig:EK_pol_dep}
\end{figure}

We note that below the threshold, despite the vertical polarisation of the excitation, we observed emission from both horizontal and vertical cavity modes due to the ultrafast intermolecular energy transfer mechanism responsible for the depolarisation of the photoluminescence (PL) in films of organic laser dyes with similar structure and properties ~\cite{Chang2007,Musser2017}. In contrast when pump power exceeds the threshold we observed sharp increase of the degree of linear polarisation with the polarisation of the lasing state inherited from the polarisation of the pump, see Fig.~\ref{fig:EK_pol_dep}.

\begin{figure}[!h]
    \centering
    \includegraphics[width=1\linewidth]{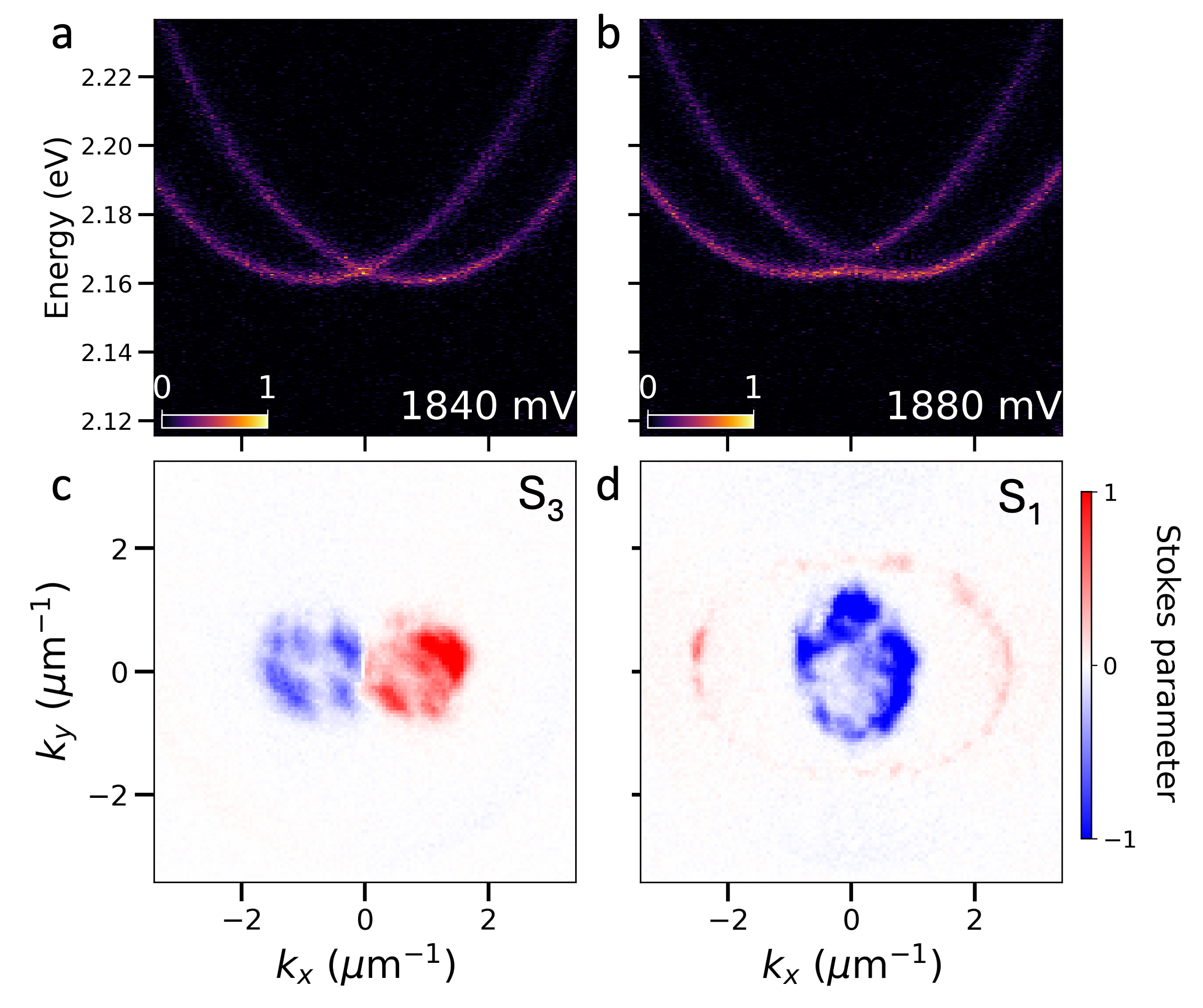}
    \caption{{\bf Emission characteristics of a single lasing state}. {\bf (a,b)} Energy-resolved momentum-space along $k_y=0$ for two values of external voltage, corresponding to {\bf (a)} 1840 mV and {\bf (b)} 1880 mV pumped below ($0.5P_{th}$) the lasing threshold and {\bf (c,d)} corresponding momentum-space distribution of the non-normalised $S_3$ {\bf (c)} and $S_1$ {\bf (1)} Stokes components of the lasing state pumped above ($1.5P_{th}$) the lasing threshold.}
    \label{fig:EK_RD}
\end{figure}

Fig.~\ref{fig:EK_RD} shows characteristics of the PL for a single isolated lasing state around Rashba-Dresselhaus (RD) regime of spin-orbit coupling (SOC). In Fig.~\ref{fig:EK_RD}(a,b) we show energy-resolved momentum space PL below the lasing threshold measured (a) exactly in RD SOC regime at 1840 mV and (b) in slightly detuned regime at 1880 mV external voltage. We further demonstrate the polarisation properties the emission in each regime providing momentum-space distribution of the non-normalised $S_3$ (Fig.~\ref{fig:EK_RD}(c)) and $S_1$ (Fig.~\ref{fig:EK_RD}(d)) Stokes components of the lasing state.

\subsection{Spatial coherence in a dyad supermode}

\begin{figure}[!h]
    \centering
    \includegraphics[width=1\linewidth]{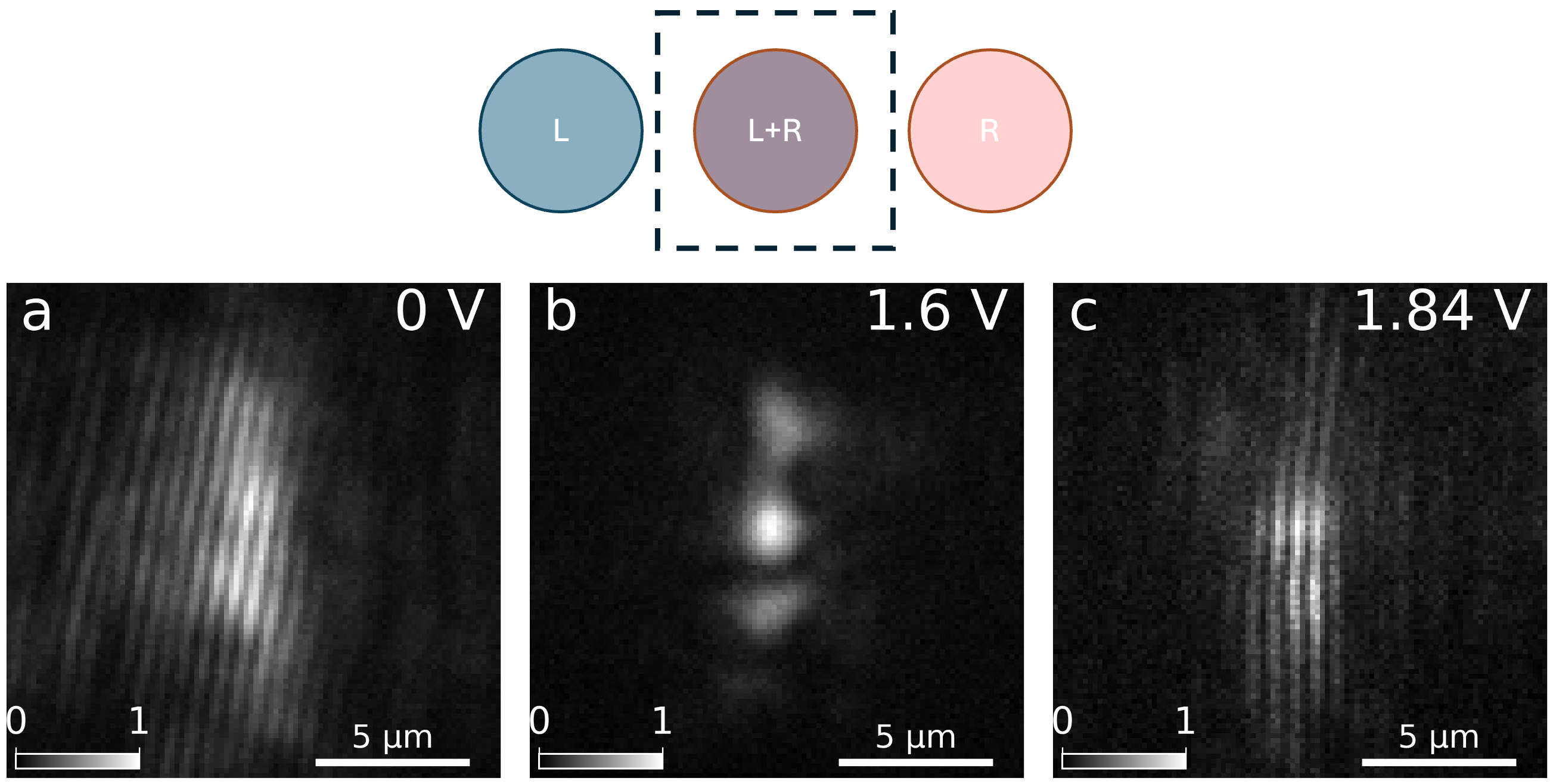}
    \caption{{\bf  Spatially extended coherence.} Experimental interferogram of the dyad supermode obtained by overlapping the real-space emission of the opposite lasing states of a dyad going through the separate arms of Mach-Zehnder interferometer at  {\bf (a)} 0~V,  {\bf (b)} 1.6~V,  {\bf (c)} 1.84~V. Schematic above represents the overlapping real-space emission of a dyad (left and right circles) from the different (red and blue) arms of the interferometer with the region depicted in images below shown by dashed square.}
    \label{fig:interf}
\end{figure}

To further support the claim of phase-locking between the spatially separated lasing states and electrical tunability of such effect we performed spatially resolved interferometry measurements for the case of dyad supermode with the spot separation distance of $13 \ \mu m$, Fig.~\ref{fig:interf}. Single shot interferometry measurements were made using Mach-Zehnder interferometer at 0 time delay to spatially overlap the real space emission from the opposite lasing states of a dyad (see schematic in Fig.~\ref{fig:interf}). We observed clear interference fringes for 0~V (Fig.~\ref{fig:interf}(a)), highlighting the synchronization between the states. Increase of the voltage to  1.6~V was accompanied by the loss of fringes visibility corresponding to desynchronisation, while further increase to 1.84~V showed revival of the fringes visibility and re-establishing of the phase-locking in a dyad. We therefore prove that the appearance of the fringe pattern in the real- and momentum-space of the emission from the array of spatially extended lasing states is accompanied by a long-range phase-coherence.

\subsection{Additional data on coupling in the presence of spin-orbit interaction}

\begin{figure}[!h]
    \centering
    \includegraphics[width=1\linewidth]{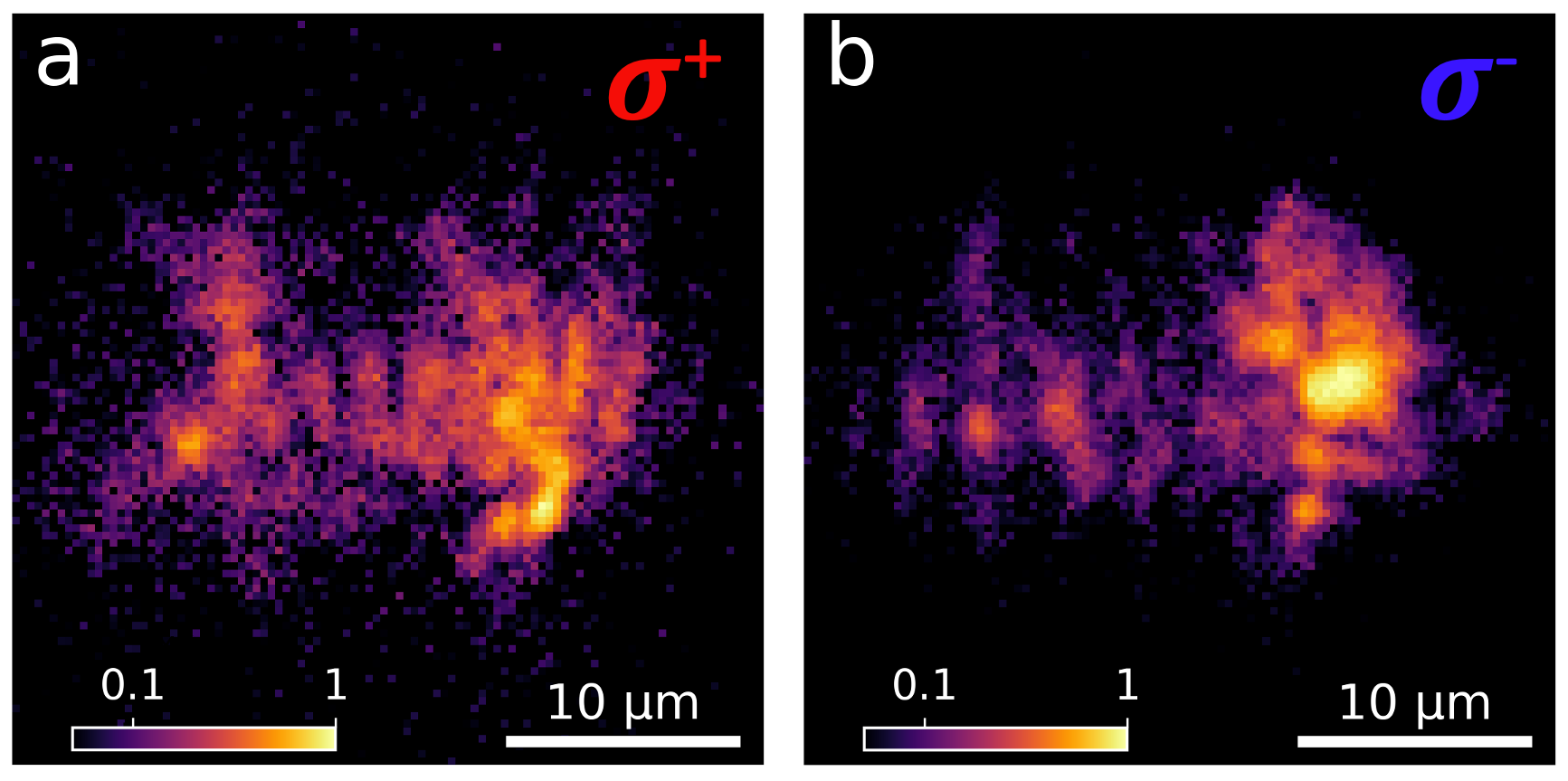}
    \caption{{\bf Dyad lasing state in the presence of spin-orbit coupling}. Right-circularly polarised {\bf (a)} and Light-circularly polarised {\bf (b)} polarised real-space distribution of the emission for two pump spots at 1840 mV showing coupled supermode lasing state. For better visibility images are illustrated in logarithmic scale.} 
    \label{fig:dyad_RD_polres}
\end{figure}

We state in Sec. 2.3 of the main manuscript that we observed the revival of the coupling between separated lasing states in a dyad configuration when system enters RD SOC regime. However, while the signature fringe pattern is clearly observed in Fourier-space, it's visibility is lowered in the real-space image of total emission due to the spatial shift between the emission patterns of opposite circular polarisation. In Fig.~\ref{fig:dyad_RD_polres} we show the corresponding experimentally measured polarisation-resolved real-space distribution of the emission from a coupled dyad state in RD SOC regime at 1840 mV external voltage. Characteristic fringe pattern is observed for both polarisation components.

\subsection{Additional data on unconventional coupling}

\begin{figure}[!h]
    \centering
    \includegraphics[width=1\linewidth]{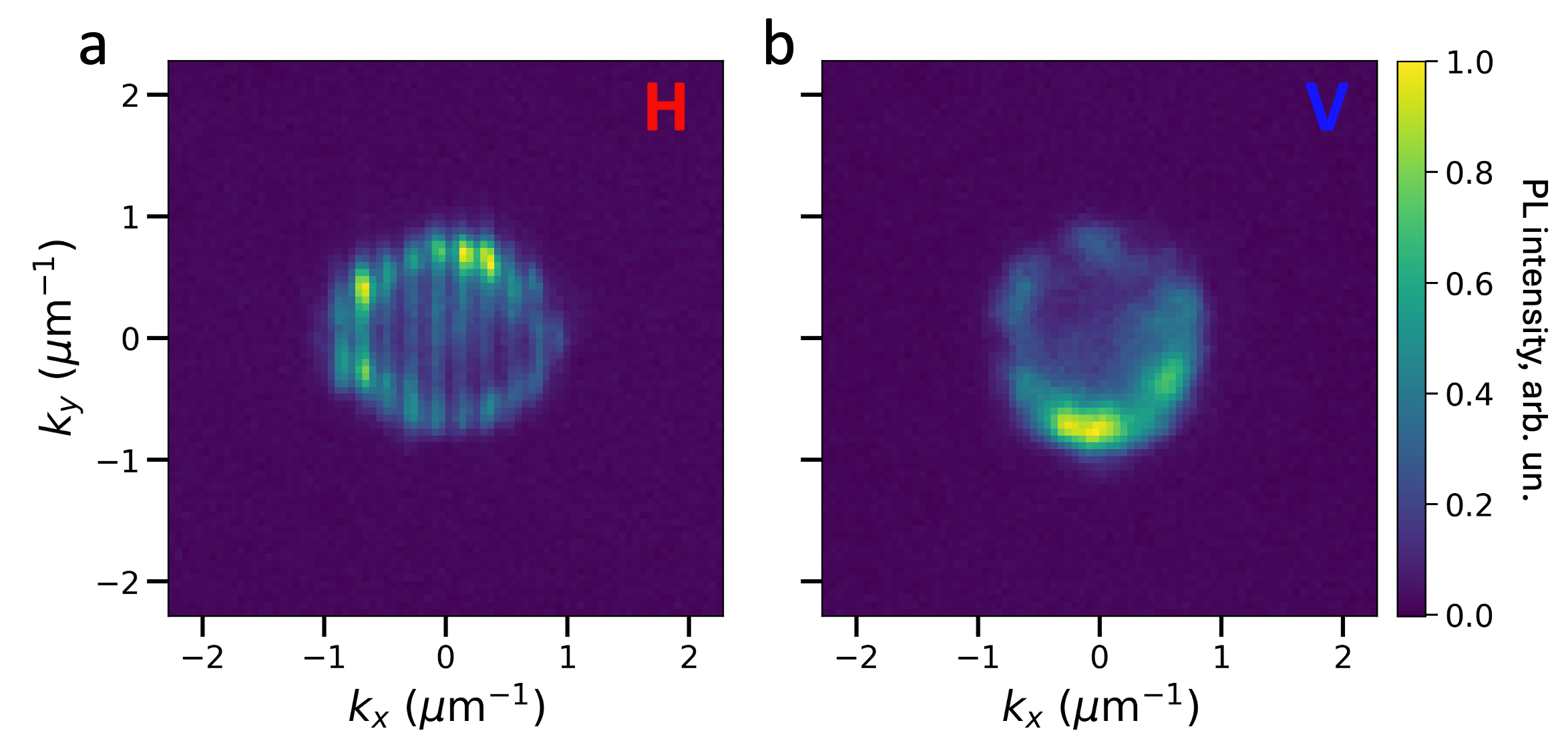}
    \caption{{\bf Next-nearest-neighbour coupling regime.}. Horisontally {\bf (a)} and vertically {\bf (b)} polarised momentum-space distribution of the emission from 3 lasing states in the NNN coupling regime.}
    \label{fig:KK_S1_triad}
\end{figure}

In Sec. 2.4 of the manuscript we demonstrate realisation of next-nearest-neighbour (NNN) coupling regime by analysing the periodicity of interference fringes visible in momentum space. In Fig.~\ref{fig:KK_S1_triad} we show polarisation resolved PL in momentum space for NNN regime of interaction corresponding to Fig.6(f) of the main manuscript. Phase-locking is manifested between horisontally polarised states, while no interference fringes are observed in vertically polarised emission.

\section{Determination of the coupling by extracting photonic component}

\begin{figure}[!h]
    \centering
    \includegraphics[width=1\linewidth]{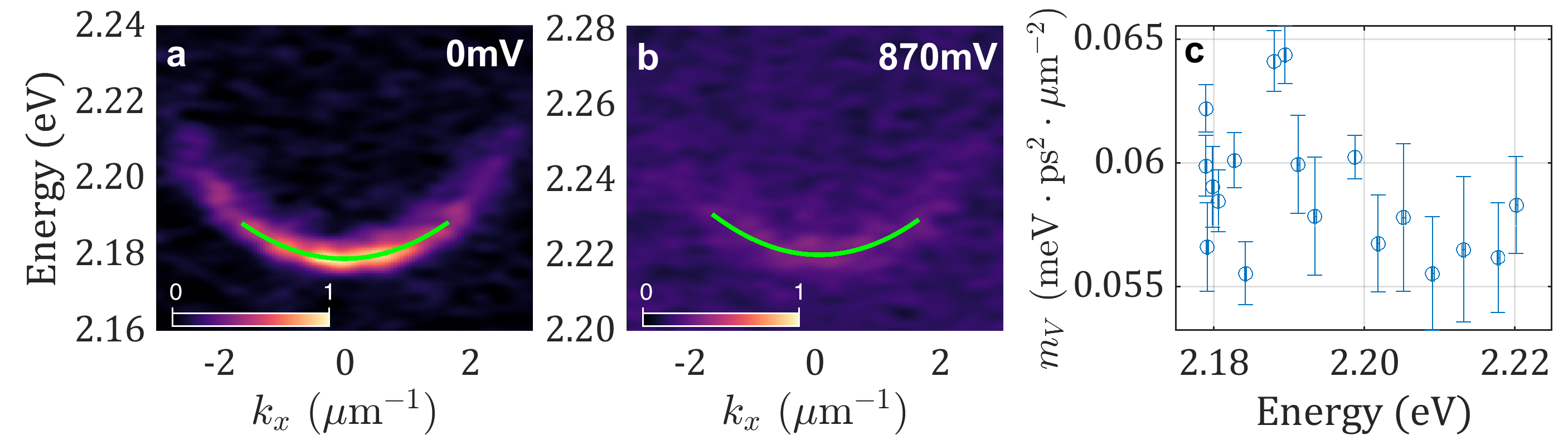}
    \caption{Experimental dispersion of the tunable vertically-polarized cavity mode below the lasing threshold for the voltage of (a) 0 mV and (b) 870 mV; the green line demonstrates the fitting by a parabola; (c) extracted mode mass for different voltages by fitting with a parabolic function; error bars include only a fitting error and do not include experimental uncertainties; the mode mass does not increase while approaching the excitonic resonance, therefore, the mode is in the weak coupling regime.} 
    \label{fig:mass_extr}
\end{figure}

The standard way of strong coupling determination is the observation of cavity mode anticrossing while the photonic mode approaches the exciton. In our sample any photonic mode has a parabolic shape and does not demonstrate anticrossing. More precisely, any mode vanishes before we start seeing anticrossing both at the high wavevector or at the low wavevector when it is shifted towards the exciton, which is applicable for tunable vertically polarized modes. However, as it has been shown in a recent study~\cite{muszynski2024observation}, when the exciton resonance is widely inhomogeneously broadened, the absence of anticrossing does not signify the absence of strong coupling for high exciton-photon detunings. A more detailed analysis of modes is needed which we perform in the following.

We follow a tunable vertically polarized photonic mode for different values of applied voltage until it vanishes due to the proximity to the exciton located at 2.4 eV for the dye P580. For each value of voltage, we fit the mode with a parabola, as shown in Fig.~\ref{fig:mass_extr}(a,b) for two values of voltage (all dispersions for voltage values in-between demonstrate a good fit by parabola as well with coefficient of determination $R^2 > 0.96$). We extract the effective mass and the bottom (mode) energy of each parabola and plot them together in Fig.~\ref{fig:mass_extr}(c). In the case of the strong coupling, this dependence has to show a gradual increase of the mode mass when approaching the exciton~\cite{muszynski2024observation}. In our case, we see that the mass remains approximately constant. Therefore, we conclude that the sample is in the weak coupling regime.

\section{Additional Simulations using the Non-Hermitian 2D Schrödinger Model}
\begin{figure}[!h]
    \centering
    \includegraphics[width=1\linewidth]{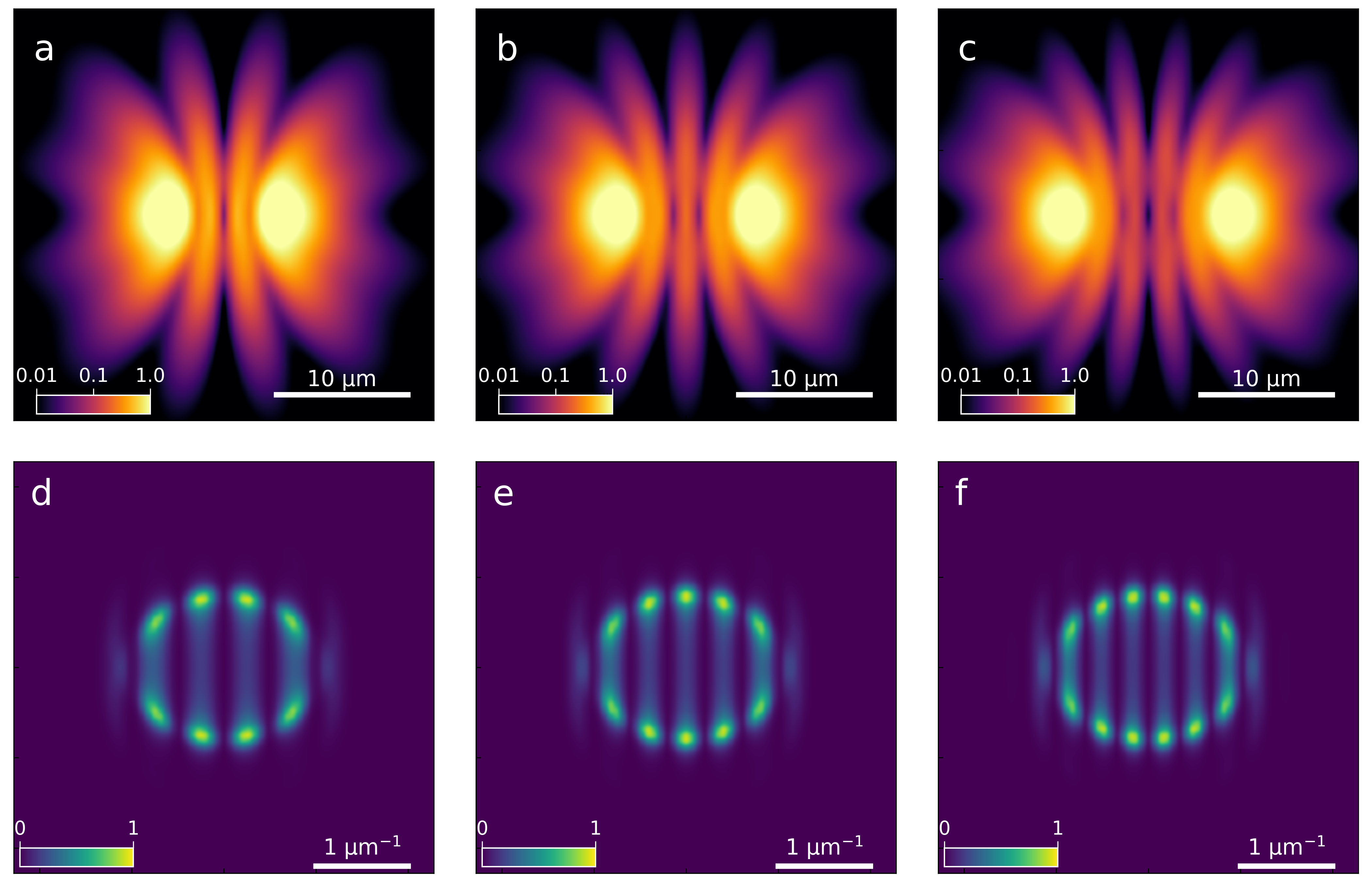}
    \caption{{\bf Simulation for the dyad supermode lasing state} for distinct distances $d$ between pump spots, corresponding to profiles shown in Fig.~2 of the main text. Panels \textbf{(a,d)} exhibit real and Fourier-space profiles for $d = \SI{9}{\micro \meter}$, while \textbf{(b,e)} and \textbf{(d,f)} correspond to $d = \SI{11}{\micro \meter}$ and $d = \SI{13}{\micro \meter}$, respectively. } 
    \label{fig:dyad_dist_simulation}
\end{figure}

\begin{figure}[!h]
    \centering
    \includegraphics[width=1\linewidth]{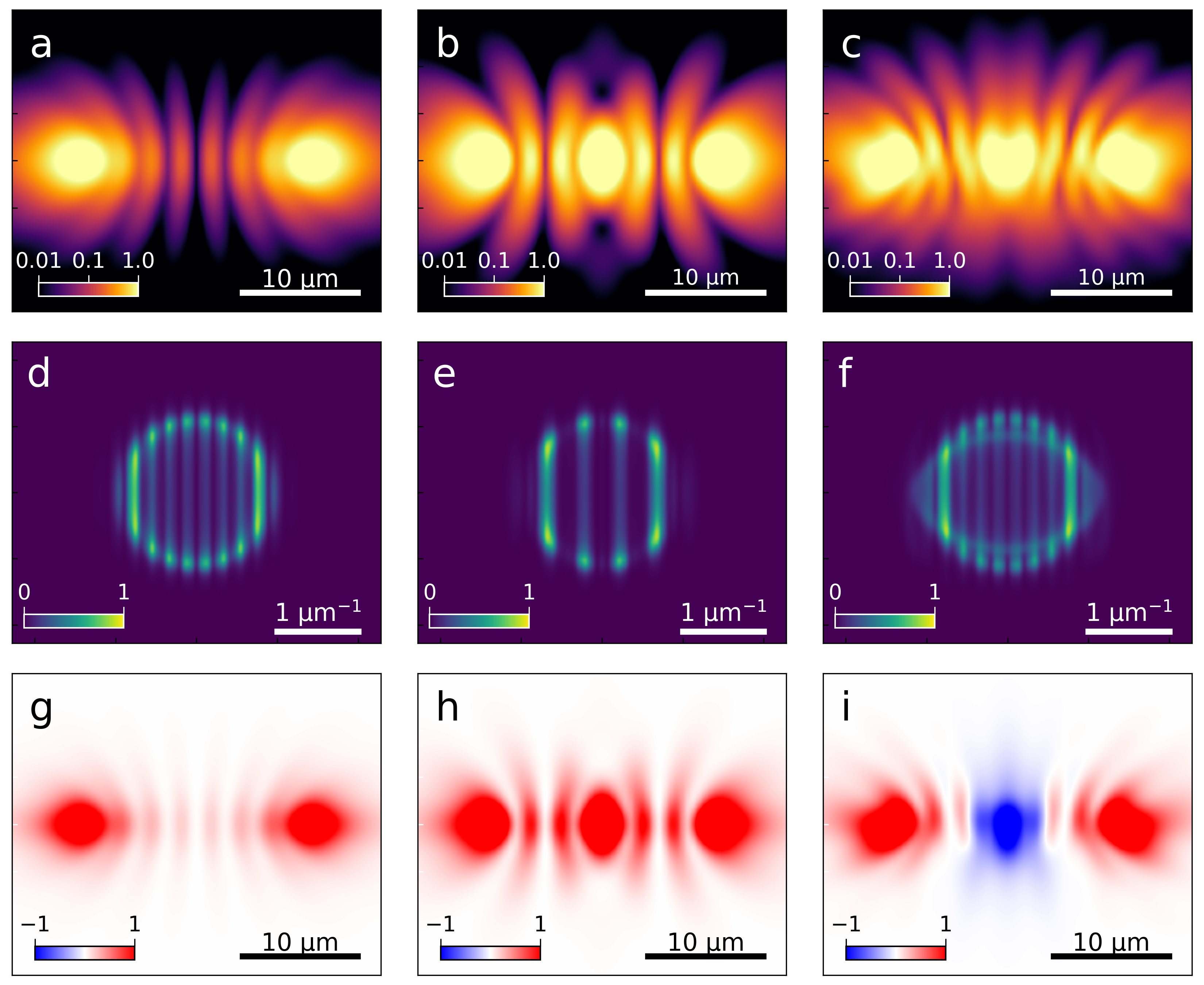}
    \caption{{\bf Simulation for an unconventional coupling in a 1D chain of pump spots} corresponding to emission profiles shown in Fig.~6 of the main text. The top and middle panels show the numerical simulations for both real and Fourier-space profiles, with the bottom panels exhibiting the real-space $S_1(\boldsymbol{r})$ component of the Stokes vector describing the degree of linear polarisation emission profile as defined in Eq.~(3) of the main text. \textbf{(a,d,g)} describes two vertically polarised pump spots at $\SI{20}{\micro \meter}$ separation distance. \textbf{(b,e,h)} show the simulations for three vertically polarised pump spots separated by $\SI{10}{\micro \meter}$ from each other. \textbf{(c,f,i)} also numerically describes three pump spots separated by $\SI{10}{\micro \meter}$, but with the central spot pumped with horizontally polarised light. } 
    \label{fig:triad_simulation}
\end{figure}

This section presents additional data for simulated real and Fourier-space distributions using the Non-Hermitian 2D Schrödinger equations (NHSE) as described in section 4.3 of the main text. Particularly,  in Fig. ~\ref{fig:dyad_dist_simulation}  we show the simulations for the dyad supermode lasing states as a function of the distance between pump spots, corresponding to cases depicted in Fig.~2 of the main text. We also depict numerical simulations for the case of unconventional coupling in the 1D chain of pump spots as seen in Fig.~\ref{fig:triad_simulation}, corresponding to experimental profiles from Fig.~6 in the main text.

\bibliography{bibliography}